\documentclass[12pt,document,nofootinbib,superscriptaddress,onecolumn,preprintnumbers,balancelastpage]{article}
\pdfoutput=1

\usepackage{jheppub_1}

\usepackage{subfig}
\usepackage[countmax]{subfloat}
\usepackage[framemethod=TikZ]{mdframed}
\usepackage{mathtools}
\usepackage{multirow}

\usepackage{slashed}

\usepackage{booktabs} 

\usepackage{array}
 
\makeatletter  
\newcommand{\thickhline}{%
    \noalign {\ifnum 0=`}\fi \hrule height 1pt
    \futurelet \reserved@a \@xhline
}

\newcolumntype{"}{@{\hskip\tabcolsep\vrule width 1pt\hskip\tabcolsep}}

\makeatother

\usepackage{latexsym}
\usepackage{amssymb}
\usepackage{epsfig,amsmath,graphics}
\usepackage{listings}
\usepackage{color}
\usepackage{dcolumn}
\usepackage{slashed}
\usepackage{cancel}
\usepackage{comment,latexsym} 
\usepackage{bm}
\usepackage{verbatim}
\usepackage{tabularx}
\usepackage{dcolumn}
\definecolor{Mahogany}{rgb}{0.62,0.24,0.15}
\definecolor{colorLink}{rgb}{0.7,0,0}
\definecolor{colorCite}{rgb}{0,.7,0}
\definecolor{colorURL}{rgb}{0,0,0.7}
\usepackage[colorlinks=true,linktocpage=true,linkcolor=black,citecolor=colorCite,urlcolor=colorURL]{hyperref}
\usepackage{setspace}
\usepackage{xspace}
\usepackage{multirow}
\usepackage{titlesec}
\usepackage[bbgreekl]{mathbbol}
\usepackage{bm}
\usepackage{float}
\usepackage{placeins}
\usepackage{afterpage}
\usepackage{footmisc}
\usepackage{breakcites}
\usepackage[most]{tcolorbox}
\tcbset{highlight math style={enhanced,colframe=green!30!black,colback=white,arc=5pt,boxrule=1.5pt}}

\usepackage[toc,page]{appendix}

\usepackage{etoolbox}
\appto\appendix{\addtocontents{toc}{\protect\setcounter{tocdepth}{1}}}

\usepackage{scalerel}

\usepackage{arydshln}

\allowdisplaybreaks

\expandafter\def\expandafter\normalsize\expandafter{%
    \normalsize
    \setlength\abovedisplayskip{8pt}
    \setlength\belowdisplayskip{8pt}
    \setlength\abovedisplayshortskip{8pt}
    \setlength\belowdisplayshortskip{8pt}
}

\newcommand{\rpii}{{\kappa_\text{I}}}
\newcommand{\rpiiii}{{\kappa_\text{III}}}

\definecolor{darkblue}{rgb}{0,0,0.5}

\definecolor{colorTC}{rgb}{.2,.7,.2}

\newcommand{\s}{\hspace{0.8pt}}

\newcommand{\qq}{\mathcal{Q}} 

\newcommand{\PP}{\mathbb{d}}

\newcommand{\RPIi}{\,\,\xrightarrow[\hspace{5pt}\text{RPI-I}\hspace{5pt}]{}\,\,}
\newcommand{\RPIii}{\,\,\xrightarrow[\hspace{5pt}\text{RPI-II}\hspace{5pt}]{}\,\,}
\newcommand{\RPIiii}{\,\,\xrightarrow[\hspace{5pt}\text{RPI-III}\hspace{5pt}]{}\,\,}
\newcommand{\gauge}{\,\,\xrightarrow[\hspace{5pt}\text{Gauge}\hspace{5pt}]{}\,\,}
\newcommand{\susy}{\,\,\xrightarrow[\hspace{5pt}\text{SUSY}\hspace{5pt}]{}\,\,}

\DeclareRobustCommand{\Sec}[1]{Sec.~\ref{#1}}
\DeclareRobustCommand{\Secs}[2]{Secs.~\ref{#1} and \ref{#2}}
\DeclareRobustCommand{\App}[1]{App.~\ref{#1}}
\DeclareRobustCommand{\Tab}[1]{Table~\ref{#1}}

\DeclareRobustCommand{\Eq}[1]{Eq.~(\ref{#1})}
\DeclareRobustCommand{\Eqs}[2]{Eqs.~(\ref{#1}) and (\ref{#2})}
\DeclareRobustCommand{\Ref}[1]{Ref.~\cite{#1}}
\DeclareRobustCommand{\Refs}[1]{Refs.~\cite{#1}}

\newcommand{\D}{\mathbb{D}}
\newcommand{\Dbar}{\bar{\mathbb{D}}}
\newcommand{\Q}{\mathbb{Q}}
\newcommand{\RCA}{\boldsymbol{\Omega}}

\newcommand{\be}{\begin{equation}}
\newcommand{\ee}{\end{equation}}
\newcommand{\bea}{\begin{eqnarray}}
\newcommand{\eea}{\end{eqnarray}}

\newcommand{\alc}{\mathcal{A}}

\newcommand{\nbp}{\bar{n}\cdot \partial} 
\newcommand{\np}{n\cdot \partial}

\newcommand{\bPhiA}{\bold \Phi_{\alc}}

\newcommand{\bPhi}{\boldsymbol{\Phi}}
\newcommand{\bPsi}{\boldsymbol{\Psi}}

\newcommand{\bV}{\boldsymbol{V}}
\newcommand{\uu}{\tilde u}

\newcommand{\bM}{ \boldsymbol M}

\newcommand{\bPhialc}{ \bold \Phi_\alc}

\preprint{MIT--CTP 5076}

\title{
Navigating Collinear Superspace
}

\author[a]{Timothy Cohen,}
\author[b]{Gilly Elor,}
\author[c]{Andrew J.~Larkoski,}
\author[d,e]{and Jesse Thaler\,}

\affiliation[a]{\footnotesize Institute of Theoretical Science, University of Oregon, Eugene, OR 97403, U.S.A.}
\affiliation[b]{\footnotesize Department of Physics, Box 1560, University of Washington, Seattle, WA 98195, U.S.A.}
\affiliation[c]{\footnotesize Physics Department, Reed College, Portland, OR 97202, U.S.A.}
\affiliation[d]{\footnotesize Center for Theoretical Physics, Massachusetts Institute of Technology, Cambridge, MA 02139, U.S.A.}
\affiliation[e]{\footnotesize Department of Physics, Harvard University, 17 Oxford Street, Cambridge, MA 02138, U.S.A.}

\emailAdd{tcohen@uoregon.edu}
\emailAdd{gelor@uw.edu}
\emailAdd{larkoski@reed.edu}
\emailAdd{jthaler@mit.edu}

\abstract{
We introduce a new set of effective field theory rules for constructing Lagrangians with $\mathcal{N} = 1$ supersymmetry in collinear superspace.
In the standard superspace treatment, superfields are functions of the coordinates $\big(x^\mu,\theta^\alpha, \theta^{\dagger \dot{\alpha}}\big)$, and supersymmetry preservation is manifest at the Lagrangian level in part due to the inclusion of auxiliary $F$- and $D$-term components.
By contrast, collinear superspace depends on a smaller set of coordinates $\big(x^\mu,\eta,\eta^\dagger\big)$, where $\eta$ is a complex Grassmann number without a spinor index.
This provides a formulation of supersymmetric theories  that depends exclusively on propagating degrees of freedom, at the expense of obscuring Lorentz invariance and introducing inverse momentum scales.
After establishing the general framework, we construct collinear superspace Lagrangians for free chiral matter and non-Abelian gauge fields.
For the latter construction, an important ingredient is a superfield representation that is simultaneously chiral, anti-chiral, and real; this novel object encodes residual gauge transformations on the light cone.
Additionally, we discuss a fundamental obstruction to constructing interacting theories with chiral matter; overcoming these issues is the subject of our companion paper, where we introduce a larger set of superfields to realize the full range of interactions compatible with $\mathcal{N} = 1$.
Along the way, we provide a novel framing of reparametrization invariance using a spinor decomposition, which provides insight into this important light-cone symmetry.
}

\begin{document} 
\maketitle

\setcounter{page}{2}
\begin{spacing}{1.2}

\pagebreak

\section{Casting Off}
\label{sec:intro}
Supersymmetry (SUSY) is a powerful tool for exploring formal aspects of field theory, including Seiberg duality~\cite{Seiberg:1994pq}, Seiberg-Witten~\cite{Seiberg:1994aj,Seiberg:1994rs}, AdS/CFT for $\mathcal{N}=4$ SUSY~\cite{Maldacena:1997re}, supersymmetric localization~\cite{Pestun:2007rz, Erickson:2000af}, on-shell recursion applied to SUSY theories~\cite{Bianchi:2008pu}, and more \cite{Komargodski:2009rz,Festuccia:2011ws,Kahn:2015mla,Ferrara:2015tyn,Dall'Agata:2016yof,Delacretaz:2016nhw,Cacciatori:2017qyd}.
Therefore, new formulations of SUSY are of great interest in their own right, especially when they can expose new formal features.
Theories with $\mathcal{N}=1$ SUSY can be expressed in superspace~\cite{Salam:1974yz,Ferrara:1974ac}, which makes SUSY manifest at the Lagrangian level by relying on non-propagating field content (including the auxiliary $F$- and $D$-terms).
A conventional formalism for systematically extending the superspace formalism to theories with $\mathcal{N} > 1$ SUSY is not known, in part because of complications associated with a proliferation of auxiliary fields, although progress has been made in harmonic and projective superspaces~\cite{Galperin:2001uw, Galperin:1985ec,Akulov:1988tm,Howe:1998jw,Davgadorj:2017ezp,Galperin:1984av,Sokatchev:1985tc,Ivanov:1984ut,Ohta:1985au,Ivanov:2003nk}.

There do exist superspace formulations that involve only propagating physical degrees of freedom, specifically in $\mathcal{N}=4$ SUSY~\cite{Mandelstam:1982cb, Brink:1982pd, Brink:1982wv}.
These constructions, however, are typically discovered by starting with a component Lagrangian and then guessing a superspace formulation that reproduces the component-level result.  
Ideally, one would want a set of effective field theory (EFT) rules for how to put together the strictly propagating degrees of freedom into superspace Lagrangians such that SUSY would be made manifest.
In this paper, we realize this goal for $\mathcal{N} = 1$ SUSY theories that do not require non-propagating $F$- and $D$-term auxiliary fields to model their interactions: free chiral matter and (non-)Abelian gauge theories.
A companion paper will provide the necessary formalism to realize theories with non-zero $F$- and $D$-terms, such as Wess-Zumino models and gauge theories with chiral matter~\cite{Cohen:2019gsc}.

Progress towards an on-shell EFT for SUSY was recently made in \Refs{Cohen:2016jzp, Cohen:2016dcl}, where the interplay of SUSY with Soft-Collinear Effective Theory (SCET)~\cite{Bauer:2000yr, Bauer:2001ct, Bauer:2001yt} was studied.
By introducing the formulation of ``collinear superspace,'' these papers arrived at a ``SUSY SCET" Lagrangian with only light-cone degrees of freedom.\footnote{Collinear superspace is closely related to light-cone superspace, which has a long history.
It was famously utilized to prove the UV finiteness of $\mathcal{N} = 4$ SYM \cite{Mandelstam:1982cb, Brink:1982pd, Brink:1982wv}.
Additional work has illustrated the utility of formulating various SUSY theories \cite{Belitsky:2004yg} (and even supergravity \cite{Kallosh:2009db}) on the light cone and more \cite{Green:1996um,Maldacena:1997re,Belitsky:2004yg,Kallosh:2009db,Hearin:2010dw,Ramond:2009hb}.
Notably, much of the original light-cone superspace literature was focused on representation theory and implications for extended objects (see, e.g., \cite{Siegel:1981ec,Brink:1981nb}), whereas the focus of our present work on collinear superspace is identifying the bottom-up rules to construct EFTs.
}
The logic used in \Refs{Cohen:2016jzp, Cohen:2016dcl} was decidedly \emph{top down}: start with a full-theory Lagrangian, integrate out non-propagating degrees of freedom directly in superspace, and truncate to leading power.
While this was a useful first step (since it is non-trivial to show that SUSY and SCET can be compatible), any self-respecting effective field theorist would only be satisfied by a fully \emph{bottom up} treatment: specify the building blocks, define their power countings and transformation properties under the relevant symmetries, and construct the (sub)leading-order Lagrangian directly, all without appealing to an underlying full theory.

In this work, we present a concrete set of rules to construct $\mathcal{N}=1$ SUSY Lagrangians directly in collinear superspace.
The key insight is to make only a subgroup of $\mathcal{N}=1$ SUSY manifest and to replace full Lorentz symmetry with reparametrization invariance (RPI) \cite{Manohar:2002fd, Becher:2014oda}.
Note that without any loss of information, the ordinary superspace coordinate $\theta^\alpha$ can be expressed as
\begin{align}
\label{eq:theta_as_eta_tildeeta}
\theta^\alpha = \xi^\alpha\s \eta + {\tilde{\xi}}^{\alpha} \s\tilde{\eta}\,,
\end{align}
where $\xi^\alpha$ and $\tilde{\xi}^\alpha$ are orthogonal commuting basis spinors that satisfy $\xi^\alpha \s\tilde{\xi}_\alpha = 1$, and $\eta$ and $\tilde{\eta}$ are complex Grassmann numbers.
Then, to reduce to collinear superspace, we simply set
\begin{align}
\label{eq:tilde_eta_zero}
\tilde{\eta} = 0 \quad \Longrightarrow \quad \theta^\alpha = \xi^\alpha\s \eta\,,
\end{align}
which halves the number of manifest supercharges.
By exploiting the RPI freedom to rotate $\xi^\alpha$ and $\eta$, we will show that this construction preserves enough Lorentz invariance to maintain the full $\mathcal{N}=1$ SUSY at the $S$-matrix level.\footnote{The restriction in \Eq{eq:tilde_eta_zero} is reminiscent of on-shell superspace \cite{Gates:1982an, Brink:1980cb, Elvang:2013cua}, with the important distinction that our construction does not require the component fields to be exactly on-shell, \emph{i.e.}, $p^2 = m^2$ is not enforced.}

With the replacement in \Eq{eq:tilde_eta_zero}, the superspace coordinate now has the unfamiliar property that $\theta^\alpha \theta_{\alpha} = \theta^\dagger_{\dot{\alpha}} \theta^{\dagger \dot{\alpha}} = 0$.
This means that one cannot include $F$- and $D$-term components in a superfield, at least not in the standard way, nor can one include non-propagating components of a  spin-1/2 matter field.
Therefore, if a self-consistent theory of collinear superspace exists with standard superfields, it must only involve propagating degrees of freedom.
We will show that this is indeed the case, and the choice in \Eq{eq:tilde_eta_zero} corresponds to expressing the theory with respect to a light-like direction $n^\mu = \tilde{\xi}\s \sigma^\mu\s \tilde{\xi}^\dagger$.
The choice of $n^\mu$ corresponds to an explicit breaking of Lorentz invariance, leading to a set of low-energy RPI constraints.  For example, the following rescaling
\begin{align}
\xi^\alpha \,\,\rightarrow\,\, e^{- \kappa/2}\, \xi^\alpha\,, \qquad \eta \,\,\rightarrow\,\, e^{\,\kappa/2}\, \eta\,,
\end{align}
is known as RPI-III, which acts like an (imaginary) internal $R$-symmetry that leaves $\theta^\alpha$ unchanged.
By imposing collinear SUSY, RPI, and simple power counting based on mass dimension, we can construct the unique gauge-invariant EFT of free chiral superfields and (non-)Abelian vector superfields at leading power.

Another set of RPI transformations, known as RPI-II, acts to rotate $\tilde{\eta}$ into $\eta$,  which is clearly incompatible with the projection in \Eq{eq:tilde_eta_zero}.
In order to have a fully Lorentz-invariant theory, however, RPI-II must also be preserved.
Because RPI-II transforms out of the collinear SUSY algebra, we can only test RPI-II on component fields, not directly on superfields.%
\footnote{An analogous situation arises when writing $\mathcal{N}=2$ Lagrangians in $\mathcal{N}=1$ superspace.  While this makes the $\mathcal{N}=1$ SUSY subgroup manifest, the full $\mathcal{N}=2$ algebra can only be tested on components.}
For the constructions in this paper, we find that RPI-II is an accidental symmetry that is only respected by  interactions that are leading-order in mass dimension. 
By the Haag-\L opusza\'nski-Sohnius extension~\cite{Haag:1974qh} of the Coleman-Mandula theorem~\cite{Coleman:1967ad}, this implies that these leading-ordering constructions exhibit the full $\mathcal{N}=1$ SUSY for the $S$-matrix, even though only collinear SUSY is manifest at the Lagrangian level.
Establishing RPI-II for higher-order terms is more subtle, though, with complications arising when one tries to use only standard superfields. 
In the companion paper~\cite{Cohen:2019gsc}, we introduce novel superfields which have non-trivial RPI transformation rules and which incorporate $F$- and $D$-term components, allowing for a description of the full range of $\mathcal{N} = 1$ interactions.

Though some of the discussion here is just a bottom-up recapitulation of the top-down physics already in \Refs{Cohen:2016jzp, Cohen:2016dcl} (with an emphasis on RPI in collinear superspace), there is a crucial new ingredient.
Gauge theories in collinear superspace are most naturally expressed in light-cone gauge with $\bar{n} \cdot A = 0$.
Without a full gauge symmetry, there seemed to be no easy way to constrain the EFT interactions to ensure gauge invariance without appealing to the full-theory Lagrangian.
As we will describe below, the light-cone gauge condition leaves a residual gauge redundancy.
Remarkably, this is encoded in a new type of superfield that is simultaneously chiral, anti-chiral, and real:
\vspace{-3mm}
\begin{align}
\D\s \RCA = 0, \qquad \bar{\D}\s \RCA = 0, \qquad \RCA = \RCA^\dagger,
\end{align}
where $\D$ and $\bar{\D}$ are covariant derivatives (without spinor indices) in collinear superspace, see \Eq{eq:defDDBar} below.
In ordinary superspace, such a field would just be a constant; in collinear superspace, this field is only constant along the light cone.
A residual gauge transformation encoded by $\RCA$ is sufficient to enforce gauge invariance for both the Abelian and non-Abelian cases.
It is intriguing to speculate that a similar object could help illuminate the structure of light-cone supergravity.

The main result of this work is to show that -- given transformation rules governed by RPI-I, RPI-III, collinear SUSY, and residual gauge redundancy -- it is possible to construct an interesting subset of collinear SUSY theories, namely those whose interactions do not require non-propagating auxiliary degrees of freedom.
We will show that RPI-II is obscured by choosing a fixed light cone to define collinear superspace, though we then go on to verify that RPI-II does not yield any useful constraints on the theories studied here, at least for the leading-order interactions.
Foreshadowing, RPI-II will impose non-trivial constraints in our companion paper~\cite{Cohen:2019gsc}, which deals with interacting theories that require the reintroduction of the non-propagating degrees of freedom.

The rest of this paper is organized as follows.
In \Sec{sec:formalism}, we introduce our formalism for constructing an on-shell superspace organized around \Eq{eq:tilde_eta_zero}, and we discuss the SUSY charges, transformations, and multiplets that manifest in such a constrained setup.
Next, we show how these ingredients transform under RPI in \Sec{sec:RPI}.
In \Sec{sec:BuldingL}, we show that the kinetic term for a chiral multiplet in collinear superspace is unique.
\Sec{sec:Gaugetheory} then applies analogous logic to Abelian and non-Abelian gauge theories.
Finally, \Sec{sec:Outlook} provides an outlook.
A more technical discussion of RPI is provided in an appendix, followed by an appendix summarizing some useful formulae.

\section{Charting Collinear Superspace}
\label{sec:formalism}
Our goal is to define a reduced $\mathcal{N} = 1$ collinear superspace which eliminates non-propagating degrees of freedom from the Lagrangian.
This construction will make heavy use of light-cone projections with spinors, allowing us to consistently remove half of superspace.
We can then construct collinear superfields that only involve the complex Grassmann coordinate $\eta$.\footnote{In all that follows, we use the mostly minus metric, the two-component spinors conventions of~\Ref{Dreiner:2008tw}, and  SUSY conventions defined in pages 449--453 of~\Ref{Binetruy:2006ad}.}

\subsection{The Light Cone in Spinor-Helicity Formalism}
\label{subsec:spinor_proj}

To define standard light-cone coordinates, one introduces two light-like directions $n^\mu$ and $\bar{n}^\mu$ which satisfy $n \cdot \bar{n} = 2$.
We then perform a spinor-helicity decomposition in terms of two bosonic spinors $\xi^\alpha$ and $\tilde{\xi}^\alpha$\,: 
\begin{equation}
\label{eq:twospinorsn_alt}
n_{\alpha \dot{\alpha}} \equiv \left( \frac{n\cdot \sigma}{2} \right)_{\alpha \dot{\alpha}} = \tilde{\xi}_\alpha\, \tilde{\xi}^{\dagger}_{\dot{\alpha}}\,, \qquad \bar{n}_{\alpha \dot{\alpha}} \equiv \left( \frac{\bar{n}\cdot \sigma}{2} \right)_{\alpha \dot{\alpha}} = \xi_\alpha\, \xi^{\dagger}_{\dot{\alpha}}\,,
\end{equation}
or, equivalently:%
\footnote{Throughout this work, we suppress spinor indices when the structure is obvious, and when no confusion with scalars can arise.}
\begin{align}
\label{eq:twospinorsn}
n^\mu = \tilde{\xi}^\dagger\s \bar{\sigma}^\mu\s \tilde{\xi} = \tilde{\xi} \s\sigma^\mu\s \tilde{\xi}^\dagger, \quad\qquad \bar{n}^\mu = \xi^\dagger\s \bar{\sigma}^\mu\s \xi  = \xi\s \sigma^\mu\s \xi^\dagger\,.
\end{align}
Because the spinors are bosonic, they satisfy
\begin{equation}
\xi^\alpha\s \xi_\alpha = \tilde{\xi}^\alpha \tilde{\xi}_\alpha = 0\,.
\end{equation}
We choose the normalization condition
\begin{equation}
\xi^\alpha\s \tilde{\xi}_\alpha = 1\,,
\end{equation}
which ensures the desired normalization for $n^\mu$ and $\bar{n}^\mu$ via a Fierz identity:
\begin{align}
\label{eq:chiXiCondition}
n \cdot \bar{n} = n^\mu \,\bar{n}_\mu = 2 \, n^{\alpha \dot{\alpha}}\, \bar{n}_{\alpha \dot{\alpha}} = 2\,\Big(\xi \s\tilde{\xi} \Big)\Big( \tilde{\xi}^\dag\s \xi^\dag \Big) = 2\,.
\end{align}
Note that $ \tilde{\xi}^ \alpha\s \xi_\alpha = - \epsilon^{ \beta \alpha}\, \xi_\alpha\s  \tilde{\xi}_\beta  = -1$.
The standard RPI transformations correspond to all possible shifts in $n^\mu$ and $\bar{n}^\mu$ such that \Eq{eq:chiXiCondition} is maintained. We derive a version of RPI that constrains possible operators in collinear superspace in \Sec{sec:RPI}.

When it is convenient to choose an explicit reference frame, a common choice is to align $n^\mu$ and $\bar{n}^\mu$ along the $z$-direction.
This \emph{canonical frame} is specified by 
\begin{align}
n^\mu = (1,0,0,1)\,, \qquad \bar{n}^\mu = (1,0,0,-1)\,,
\end{align}
which is equivalent to the fixing the spinors to\footnote{Note that this is consistent due to the unfortunate fact that $\epsilon^{12} = - \epsilon^{21} = \epsilon_{21} = - \epsilon_{12} = 1$.}
\begin{align}
\label{eq:bosoniccanonicalspinors}
\xi^\alpha = (0,1)\,,\qquad \xi_\alpha = (-1,0)^{\intercal}, \qquad \tilde{\xi}^\alpha = (1,0)\,, \qquad \tilde{\xi}_\alpha = (0,1)^{\intercal}\,,
\end{align}
as can be verified using \Eq{eq:twospinorsn}.
As we show below, this frame choice is equivalent to working in the collinear superspace frame developed in~\Ref{Cohen:2016jzp}.

Any operator can be projected along the $\xi_\alpha$ and $\tilde{\xi}_\alpha$ spinor axes.
Consider the differential operator $\sigma^\mu\, \partial_\mu$, where $\partial_\mu$ is the four-vector partial derivative.
We can construct differential operators on the light-cone as
\begin{align}
\label{eq:Funnyd}
\PP &=  \bar{n} \cdot \partial =  \xi^\alpha (\sigma \cdot \partial)_{\alpha \dot{\alpha}} \, \xi^{\dagger \dot{\alpha}} \, ,  &
\tilde{\PP} &= n \cdot \partial  = \tilde{\xi}^\alpha (\sigma \cdot \partial)_{\alpha \dot{\alpha}} \, \tilde{\xi}^{\dagger \dot{\alpha}} \, ,\notag\\[8pt]  
\PP_\perp & =  \xi^\alpha (\sigma \cdot \partial)_{\alpha \dot{\alpha}} \, \tilde{\xi}^{\dagger \dot{\alpha}} \, , &
\PP_\perp^* & = \tilde{\xi}^\alpha (\sigma \cdot \partial)_{\alpha \dot{\alpha}} \, \xi^{\dagger \dot{\alpha}}. 
\end{align}
Here, we have introduced the $\PP$ notation to emphasize that we have \emph{not} made a specific frame choice.\footnote{Note that $\PP / \tilde{\PP}$, which are equivalent to $\nbp/ \np$, are often referred to in the literature as $\partial_{\pm}$.  See \Ref{Leibbrandt:1983pj} for a review on standard light-cone conventions. We adopt the new $\PP/\tilde{\PP}$ convention to emphasize that we can formulate the theory without appealing to a specific frame. We are unaware of any light-cone-independent analog of $\PP_\perp$ in the literature.}
The d'Alembertian can be expressed along an unspecified light-cone direction as
\begin{align}
\Box = \PP \s \tilde{\PP} - \PP_\perp^* \PP_\perp \, .
\end{align}

\subsection{Projecting Spinors and Gauge Fields}
\label{sec:ProjectGauge}

Throughout this paper, we make use of the light-cone decomposition of fields that carry Lorentz indices.
We begin by discussing the light-cone projections for a left-handed two-component Weyl spinor $u_\alpha$.
Recall that $u_\alpha$ may be decomposed (by acting with chiral projection operators) onto a helicity component that is aligned with the light cone and another that is anti-aligned.%
\footnote{This decomposition is valid for both massless and massive fermions, though in the massive case, the helicity components are not mass eigenstates.}
Specifically, we can decompose
\vspace{-3mm}
\begin{align}
\label{eq:udecom}
&u_\alpha =  \tilde{\xi}_\alpha\s u - \xi_\alpha\s \tilde{u}\,, 
\end{align}
with
\vspace{-3mm}
\begin{align}
\label{eq:udecom_reverse}
 \xi^\alpha \s u_\alpha = u \quad \text{and} \quad \tilde{\xi}^\alpha\s  u_\alpha = \tilde{u}\,.
\end{align}
Here, $u$ is the helicity component that propagates in collinear superspace, while $\uu$ is the other helicity which will play a role in~\Ref{Cohen:2019gsc}.

Next, we decompose the full Lorentz four-vector field $A^\mu$ as
\begin{align}
\left(\sigma \cdot A \right)_{\alpha \dot{\alpha}} = \xi_\alpha\, \xi^{\dagger}_{\dot{\alpha}}\, n\cdot A + \tilde{\xi}_\alpha\, \tilde{\xi}^\dagger_{\dot{\alpha}}\, \bar{n}\cdot A+\sqrt{2} \,  \xi_\alpha\, \tilde{\xi}^\dagger_{\dot{\alpha}}\, \alc^* +\sqrt{2} \,  \tilde{\xi}_\alpha\, \xi^{\dagger}_{\dot{\alpha}}\, \alc\,,
\end{align}
where we have projected the gauge field $A^\mu$ onto a complex ``light-cone gauge" scalar using
\begin{align}
\label{eq:def_script_A}
\alc =\frac{1}{\sqrt{2}} \xi^\alpha\s (\sigma \cdot A )_{\alpha \dot{\alpha}}\s \tilde{\xi}^{\dagger \dot{\alpha}} \, , \quad \quad \alc^* =  \frac{1}{\sqrt{2}} \tilde \xi^\alpha\s (\sigma \cdot A )_{\alpha \dot{\alpha}}\s \xi^{\dagger \dot{\alpha}} \,.
\end{align}
This $\alc$ field encodes the two propagating degrees of freedom of a gauge field, \emph{i.e.}, those that are transverse to the light cone. 

The two other degrees of freedom, $n\cdot A$ and $\bar{n} \cdot A$, while non-propagating (and therefore not the focus of the current work) can be obtained via the projections 
\begin{align}
\bar n \cdot A = \xi^\alpha\s (\sigma \cdot A)_{\alpha \dot{\alpha}}\s \xi^{\dot{\alpha} \dagger} \, , \quad \quad  n\cdot A = \tilde \xi^\alpha\s (\sigma \cdot A)_{\alpha \dot{\alpha}}\s \tilde \xi^{\dagger \dot{\alpha}} \,.
\label{eq:nAandnBarA}
\end{align}
The $\bar{n}\cdot A$ mode may be eliminated by enforcing light-cone gauge, as will be done in what follows.
Furthermore, it is straightforward to see that no light-cone time derivatives act on $n\cdot A$, and as such it can be treated as a non-propagating component of the gauge field.
It is therefore prudent to integrate it out using the equations of motion, which yields the well known light-cone Lagrangian for the gauge field, see \emph{e.g.}~\Ref{Leibbrandt:1983pj}.

The Lagrangians constructed in \Secs{sec:BuldingL}{sec:Gaugetheory} will involve only the propagating degrees of freedom:  $\phi$, $u$, and $\alc$.
As we will see in \Sec{subsec:RPIcomponents}, however, RPI-II transforms us away from our chosen slice of collinear superspace.
For this reason, it will often be convenient to make RPI-II manifest by introducing auxiliary degrees of freedom:  $\uu$, $n \cdot A$, and $\bar{n}\cdot A$.
We have just shown that these fields correspond to projections of the full Lorentz representations $u^\alpha$ and $A^\mu$, so we can derive their RPI properties from the ``top down'' using the Lorentz algebra (see \App{app:RPIgen}).
That said, we will construct the actual Lagrangians from the ``bottom up,'' relying on the auxiliary fields only to check for possible RPI-II constraints on the low-energy effective theory.
At the end of the day, we will find that RPI-II does not introduce any additional requirements on the theories studied here.

\subsection{Projecting Superspace Coordinates}

We can now use the light-cone spinors to isolate half of superspace.
Starting from the standard $\mathcal{N} = 1$ superspace coordinate $\theta^\alpha$, we can construct two spinor projections:%
\footnote{For later convenience, we have chosen a different sign convention for the projection of a superspace coordinate than for the projection of a spinor field in \Eq{eq:udecom}.}
\begin{align}
\label{eq:eta_def}
\begin{array}{l}\eta =  \tilde{\xi}^\alpha\, \theta_\alpha \\[5pt] \tilde{\eta} = -\xi^\alpha\, \theta_\alpha \end{array}\,, \qquad \Longleftrightarrow \qquad \theta^\alpha = \xi^\alpha\, \eta + \tilde{\xi}^{\alpha}\, \tilde{\eta}\,,
\end{align}
where $\eta$ and $\tilde{\eta}$ are complex Grassmann numbers which \emph{do not} carry a spinor index.
Note that the minus sign in \Eq{eq:eta_def} results from the identity $ \xi \s \tilde{\xi}= 1 = - \tilde{\xi} \s\xi $.
The conjugate superspace coordinates are defined analogously:  
\vspace{-1mm}
\begin{align}
\begin{array}{l}\eta^\dagger =  - \tilde{\xi}^\dagger_{\dot{\alpha}}\, \theta^{\dagger\dot{\alpha}}\\[5pt]  \tilde{\eta}^\dagger = \xi^\dagger_{\dot{\alpha}}\, \theta^{\dagger\dot{\alpha}}\end{array}\,, \qquad \Longleftrightarrow \qquad \theta^{\dagger\dot{\alpha}} = \eta^\dagger \, \xi^{\dagger \dot{\alpha}} + \tilde{\eta}^\dagger \, \tilde{\xi}^{\dagger \dot{\alpha}}\,.
\end{align}
It is helpful to note that $\big(\xi^\alpha\s \tilde{\xi}_\alpha\big)^\dagger = \big(\tilde{\xi}_\alpha\big)^\dagger \big(\xi^\alpha\big)^\dagger = \tilde{\xi}^\dagger_{\dot{\alpha}}\s \xi^{\dagger \dot{\alpha}} =1 = - \xi^\dagger_{\dot{\alpha}}\s \tilde{\xi}^{\dagger \dot{\alpha}}$. 
As expected from their anti-commuting nature, one can verify that $\eta^2 = \big(\eta^\dagger\big)^2 = \big\{\eta, \eta^\dagger \big\} = 0$.
Crucially, $\xi_\alpha$ and $ \xi^{\dagger}_{\dot{\alpha}}$ are complex conjugates of each other, such that the superfield $\boldsymbol{\Phi}^\dagger$ will be the conjugate of $\boldsymbol{\Phi}$ (see \Eqs{eq:chiralsuperfield}{eq:antichiralsuperfield} below).
We choose as convention for the mass dimension 
\begin{align}
\big[\,\xi\,\big] = 0\,, \qquad \big[\,\eta\,\big] = -1/2 \,,
\end{align}
such that the standard mass dimension $\big[\,\theta\,\big] = -1/2$ is maintained.

We can perform a similar decomposition of the supercoordinate derivative: 
\begin{align}
\frac{\partial \eta\,}{\,\partial \theta^\alpha} = \tilde{\xi}^\alpha\quad \text{and} \quad \frac{\partial \tilde{\eta}\,}{\,\partial \theta^\alpha} =  - \xi^\alpha \qquad \Longrightarrow \qquad \frac{\partial}{\partial \theta_\alpha} = \tilde{\xi}^\alpha \frac{\partial}{\partial \eta} - \xi^\alpha \frac{\partial}{\partial \tilde{\eta}}\,\,.
\end{align}
This is consistent with the anti-commutation relations:
\begin{equation}
\left\{\eta,\, \frac{\partial}{\partial \eta}  \right\} = 1\,, \qquad \left\{\tilde{\eta},\, \frac{\partial}{\partial \tilde{\eta}}  \right\} = 1\,, \qquad \left\{\eta, \,\frac{\partial}{\partial \tilde{\eta}}  \right\} = 0\,, \qquad \left\{\tilde{\eta}, \,\frac{\partial}{\partial \eta}  \right\} = 0\, . 
\end{equation}
Now that $\eta$ and $\tilde{\eta}$ are factorized, reducing to collinear superspace is as simple as
\begin{align}
\Aboxed{\begin{minipage}{0.09\linewidth}\vspace{7pt}$\hspace{7pt}\tilde{\eta} = 0$\vspace{7pt}\end{minipage}} \qquad \Longrightarrow  \qquad \theta^\alpha = \xi^\alpha\, \eta\,, \qquad \frac{\partial}{\partial \theta_\alpha} =  \tilde{\xi}^\alpha \frac{\partial}{\partial \eta}\,.
\end{align}
With this restriction, it follows that $\theta^\alpha \theta_{\alpha} = 0$, implying that the usual $F$- and $D$-term auxiliary fields must be absent in this setup (see \Sec{sec:Superfields}). 

\subsection{The Collinear SUSY Algebra}

Using this light-cone spinor decomposition, the supercharges and superspace derivatives take a simple form.
Starting from the full $\mathcal{N}=1$ SUSY algebra,%
\footnote{We use the superscript ``full" to be explicit when working with objects of the full $\mathcal{N}=1$ theory, or in situations where we have not yet restricted to collinear superspace, \emph{i.e.} set $\tilde{\eta} = 0$.} 
\begin{align}
 \Big \{ \qq_\alpha^{\rm full}, \qq^{\dagger \rm full}_{\dot{\alpha}} \Big \} =  -2\s i\, (\sigma \cdot \partial)_{\alpha \dot{\alpha}}  \, ,
\end{align}
we can construct various sub-algebras by contracting with the $\xi^\alpha$ and $\tilde{\xi}^\alpha$ spinors.
For instance, contracting with $\xi^\alpha$ and $\xi^\dagger_{\dot{\alpha}}$, we obtain
\begin{align}
\label{eq:reducedSUSYalg}
 \Big \{ \Q^{\rm full}, \Q^{\dagger \rm full} \Big \} =  - 2\s i\, \PP\,,  \qquad \text{with} \qquad \Q^{\rm full} \equiv \xi^\alpha\, \qq^{\rm full}_\alpha\,, \qquad \Q^{\dagger \rm full} \equiv \qq^{\dagger \rm full}_{\dot{\alpha}}\, \xi^{\dagger \dot{\alpha}}\,,
\end{align}
where $\Q^{\rm full}$ and $\Q^{\dagger \rm full}$ are collinearly-projected SUSY generators without spinor indices.
The collinear sub-algebra in \Eq{eq:reducedSUSYalg} will be the focus of this study.

Without loss of generality, the original SUSY generators can be expressed in terms of the $\eta$ and $\tilde{\eta}$ coordinates as  
\begin{align}
\qq_\alpha^\text{full} =i \s \tilde{\xi}_\alpha \,\frac{\partial}{\partial \eta} - i\s \xi_\alpha\, \frac{\partial}{\partial \tilde{\eta}} -  (\sigma \cdot \partial)_{\alpha \dot{\alpha}} \big(\eta^\dagger\, \xi^{\dagger \dot{\alpha}} + \tilde{\eta}^\dagger\, \tilde{\xi}^{\dagger \dot{\alpha}}\big) \, ,
\end{align}
and similarly for $\big(\qq^{ \text{full}}_{\alpha}\big)^\dagger$.
Using the definition of $\Q$ in \Eq{eq:reducedSUSYalg}, this yields
\begin{align}
\Q^\text{full} = i\s \frac{\partial}{\partial \eta}  - \eta^\dagger\, \PP - \tilde{\eta}^\dagger\, \PP_\perp\,. 
\end{align}
Note that the $\PP_\perp$ term does not contribute to the anti-commutator in \Eq{eq:reducedSUSYalg} since $\big(\Q^{ \text{full}}\big)^\dagger$ depends on $\partial/\partial \eta^\dagger$, not on $\partial/\partial \tilde{\eta}^\dagger$.

To restrict to collinear superspace, we simply set $\tilde{\eta} = 0$.
The collinear SUSY generators are now
\begin{align}
\Q \equiv \xi^\alpha \qq_\alpha^{\rm full} \bigg|_{\tilde{\eta} = 0}   = i\s \frac{\partial}{\partial \eta} -  \eta^\dagger \,\PP, \qquad \Q^\dagger \equiv \qq^{\dagger \rm full}_{\dot{\alpha}}   \,\xi^{\dagger \dot{\alpha}} \bigg|_{\tilde{\eta} = 0} = i\s \frac{\partial}{\partial \eta^\dagger} - \eta\, \PP\,.
\label{eq:defCollinearSUSYGens}
\end{align}
Even with this restriction, the collinear versions of $\Q$ and $\Q^\dagger$ still satisfy \Eq{eq:reducedSUSYalg}, \emph{i.e.}, $\big\{ \Q, \Q^{\dagger} \big \} =  - 2\s i\, \PP\,$. 
Note that $\PP$ commutes with both collinear SUSY generators: 
\begin{align}
\Big[ \PP, \Q \Big] = 0 = \Big[ \PP, \bar{\Q} \Big]\,.
\end{align}
When using the canonical frame in \Eq{eq:bosoniccanonicalspinors}, this sub-algebra is equivalent to the collinear superspace algebra in \Refs{Cohen:2016jzp, Cohen:2016dcl}, given by
\begin{align}
\Big \{\Q, \Q^\dagger \Big \}  =  \Big\{ \qq_2, \qq^\dagger_{\dot{2}} \Big\} \,, \quad \quad  \,\,\, \Big\{\qq_2, \qq^\dagger_{\dot{1}} \Big\} &=\Big\{\qq_1, \qq^\dagger_{\dot{2}} \Big\} =\Big\{\qq_1, \qq^\dagger_{\dot{1}} \Big\}  = 0\,.
\end{align}
Closure of this sub-algebra will be discussed \Sec{sec:SUSYTrans}.

For completeness, we note that other projections yield
\begin{align}
& \Big \{\xi^\alpha\, \qq_\alpha, \qq^\dagger_{\dot{\alpha}}\, \tilde{\xi}^{\dagger \dot{\alpha}} \Big \}  = - 2\s i\, \PP_\perp \, , \quad  \Big \{\tilde{\xi}^\alpha\, \qq_\alpha, \qq^\dagger_{\dot{\alpha}}\, \xi^{\dagger \dot{\alpha}} \Big \}  = - 2\s i\, \PP_\perp^* \, , \\[8pt] \nonumber
& \hspace{60pt}  \Big \{\tilde{\xi}^\alpha\, \qq_\alpha, \qq^\dagger_{\dot{\alpha}}\, \tilde{\xi}^{\dagger \dot{\alpha}} \Big \}  = - 2\s i\, \tilde{\PP} \, ,  
\end{align}
corresponding to different sub-algebras of the full $\mathcal{N} = 1$ SUSY.
In this way, the spinors $\xi^\alpha$ and $\tilde{\xi}^\alpha$ allow us to define SUSY sub-algebras that point along the collinear, anti-collinear, and transverse directions.

\subsection{Collinear Super-Covariant Derivatives}

In order to manipulate and restrict superfields, it is useful to define collinear super-covariant derivatives.
These can be obtained by projecting the ordinary super-covariant derivatives using the light-cone spinors.
Starting from the full superspace derivative
\begin{align}
&\mathcal{D}_\alpha^\text{full} = \frac{\partial }{\partial \theta_\alpha} - i \s (\sigma \cdot \partial)_{\alpha \dot{\alpha}}\s \theta^{\dagger \dot{\alpha}} =  \tilde{\xi}_\alpha \frac{\partial}{\partial \eta} - \xi_\alpha \frac{\partial}{\partial \tilde{\eta}} -i\s (\sigma \cdot \partial)_{\alpha \dot{\alpha}} \big(\eta^\dagger \,\xi^{\dagger \dot{\alpha}} + \tilde{\eta}^\dagger\, \tilde{\xi}^{\dagger \dot{\alpha}}\big) \, , 
\end{align}
we can reduce to collinear superspace operators by setting $\tilde{\eta} = 0$:
\begin{align}
\D \equiv \xi^\alpha \mathcal{D}_\alpha^\text{full}  \bigg|_{\tilde{\eta} = 0}  = \frac{\partial}{\partial \eta} - i\s \eta^\dagger\, \PP,  \qquad \bar{\D} \equiv \bar{\mathcal{D}}_{\dot{\alpha}}^\text{full} \xi^{\dagger \dot{\alpha}}  \bigg|_{\tilde{\eta} = 0}  =   \frac{\partial}{\partial \eta^\dagger} - i\s \eta\, \PP\, ,
\label{eq:defDDBar}
\end{align}
where these operators carry mass dimension $\big[\,\D\,\big] =  \big[\,\bar{\D}\,\big] = 1/2$. 
We see that
\begin{align}
\Big\{ \D, \bar{\D} \Big \}  = -2\s i\, \PP, \qquad \Big\{\D, \Q\Big\} = 0 = \Big\{\D, \Q^\dagger\Big\} = \Big\{\bar{\D}, \Q\Big\} = \Big\{\bar{\D}, \Q^\dagger\Big\},
\end{align}
so these objects behave as superspace derivatives in our constrained superspace.
In particular, $\D$ or $\bar{\D}$ acting on a collinear superfield yields another collinear superfield.

A number of properties of $\mathcal{D}_\alpha^\text{full}$ and $\bar{\mathcal{D}}_{\dot{\alpha}}^\text{full}$ carry over to $\D$ and $\bar{\D}$.
For example, one can perform integration by parts under the collinear superspace integral $\int \text{d} \eta \, \text{d}\eta^\dagger$.
One key difference, however, is that
\vspace{-2mm}
\begin{equation}
\D^2 = \bar{\D}^2 = 0\,,
\end{equation}
since we only have a single Grassmann coordinate $\eta$ after setting $\tilde{\eta} = 0$.
As usual, $\D$ and $\bar{\D}$ allow us to define a notion of chirality for a superfield, as will be discussed next. 

\subsection{Collinear Superfields}
\label{sec:Superfields}
A generic collinear superfield is any function of $\big(x^\mu,\eta,\eta^\dagger\big)$.
Here, we focus on superfields that do not carry any Lorentz indices, with the idea being that such indices could always be contracted with $\xi^\alpha$ and $\tilde{\xi}^\alpha$ to form a Lorentz scalar.%
\footnote{Superfields with non-trivial Lorentz structure will be utilized in the companion paper \cite{Cohen:2019gsc}.} 
Due to its Grassman nature, $\eta^2 = 0$, the most general bosonic scalar superfield is
\begin{align}
\label{eq:Snotation}
\boldsymbol{S}\big(x,\eta,\eta^\dagger\big) = a(x) + \eta \, b(x) + \eta^\dagger c(x) + \eta^\dagger \eta \, v(x)\,,
\end{align}
where $a$ and $v$ are complex scalar fields, $b$ and $c$ are complex Grassmann fields, and we follow the common practice of using bold font to delineate a superfield.
To make this look more familiar, we could instead take an ordinary superfield written in terms of $\theta^\alpha$, and just make the replacement $\theta^\alpha = \xi^\alpha \,\eta$, remembering that $\theta^2 = 0$.
This yields
\begin{align}
\label{eq:altSnotation}
\boldsymbol{S} = a + \eta \, \xi^\alpha\, b_\alpha + \eta^\dagger\, \xi^\dagger_{\dot{\alpha}} \,c^{\dot{\alpha}} + \eta^\dagger \eta \, \xi \big(\sigma^\mu v_\mu\big) \xi^\dagger\,,
\end{align}
where again $a$ is a complex scalar, $b_\alpha$ is a spinor, $c^{\dot{\alpha}}$ is an anti-spinor, and $v^\mu$ is a vector.
Of course, these different ways of writing $\boldsymbol{S}$ contain the exact same information, with $b \equiv \xi^\alpha \,b_\alpha$, $c  \equiv \xi^\dagger_{\dot{\alpha}} \,c^{\dot{\alpha}}$, and $v \equiv \xi \big(\sigma^\mu v_\mu\big) \xi^\dagger$.

From this generic collinear superfield, we can apply constraints in the usual way:
\begin{itemize}
\item{Chiral: $\bar{\D}\s \boldsymbol{\Phi} = 0$\,;}
\item{Anti-Chiral: $\D\s \boldsymbol{\Phi}^\dagger = 0$\,;}
\item{Real: $\boldsymbol{V} = \boldsymbol{V}^\dagger$\,.}
\end{itemize}
These are analogous to the representations in ordinary $\mathcal{N} = 1$ SUSY, with an important twist:  because $\bar{\D}^2 = 0$, acting a \emph{single} $\bar{\D}$ on any superfield gives a chiral superfield.
For the same reason, there is no notion of a linear superfield $\boldsymbol{L}$, since~$\bar{\D}^2\s \boldsymbol{L} = \D^2\s \boldsymbol{L} = 0$.

Focusing on the components of a chiral multiplet $\boldsymbol{\Phi}$,  
\begin{align}
\label{eq:chiralsuperfield}
\boldsymbol{\Phi}\big(x,\eta,\eta^\dagger\big) = \phi(x) + \sqrt{2} \, \eta \, u(x) + i \s \eta^\dagger \eta \, \PP \phi(x) \,,
\end{align}
it is clear that this representation is built from a complex scalar $\phi$ degree of freedom and a single helicity fermionic degree of freedom $u \equiv \xi^\alpha \,u_\alpha$, \emph{i.e.}\ an anti-commuting Lorentz scalar.
It is easy to check that the chirality condition is satisfied since $\bar{\D} \boldsymbol{\Phi} = i\s \eta \, \PP \phi - i\s \eta \, \PP \phi = 0$.
Similarly, an anti-chiral superfield can be written as
\begin{align}
\label{eq:antichiralsuperfield}
\bPhi^\dagger\big(x,\eta,\eta^\dagger\big) = \phi^*(x) + \sqrt{2} \, \eta^\dagger u^\dagger (x) - i \s\eta^\dagger \eta \, \PP \phi^*(x)  \, ,
\end{align}
where again $u^\dagger \equiv \xi^\dagger_{\dot{\alpha}} u^{\dagger \dot{\alpha}}$ is the propagating helicity of the fermion. 
Note that \Eq{eq:antichiralsuperfield} is indeed the complex conjugate of \Eq{eq:chiralsuperfield}.
These chiral superfields can be used as building blocks to generate additional superfields by acting on them with superspace derivatives:
\begin{align}
\label{eq:DPhicomponents}
\D\s \bPhi  &= \sqrt{2} \, u - 2 \s i\, \eta^\dagger \PP \phi - i\s \sqrt{2} \, \eta^\dagger \eta \, \PP u \,, \notag\\[5pt]
\bar{\D} \s\bPhi^\dagger  &=  \sqrt{2} \, u^\dagger - 2\s i\, \eta \, \PP \phi^* + i\s \sqrt{2} \, \eta^\dagger  \eta \, \PP u^\dagger \,.
\end{align}

Next, consider a real superfield field $\boldsymbol{V}$, written in the notation of \Eq{eq:altSnotation},
\begin{align}
\label{eq:vector}
\bV\big(x,\eta,\eta^\dagger\big) = a(x) + i\s \eta \, \xi^\alpha\, b_\alpha(x) - i\s \eta^\dagger\, \xi^\dagger_{\dot{\alpha}} \,b^{\dagger \dot{\alpha}}(x) + \eta\, \eta^\dagger \, \xi \big(\sigma \cdot v(x)\big) \xi^\dagger \, ,
\end{align}
where $a$ is a real scalar, $b_\alpha$ is a spinor, and $v^\mu$ is a real vector.
In the standard $\mathcal{N} = 1$ SUSY approach, real superfields are used to encode gauge fields and gauginos, but this is not possible in collinear superspace for a few reasons.
First, $v^\mu$ in \Eq{eq:vector} only contains one propagating degree of freedom, instead of the two helicities needed for a physical gauge field.
Second, $b_\alpha$ has the wrong mass dimension (and the wrong gauge transformation properties) to play the role of the gaugino.
Third, the usual approach to constructing the gauge field strength via $\boldsymbol{W}_\alpha = \big(\bar{\mathcal{D}}^\text{full}\big)^2 \mathcal{D}_\alpha^\text{full}\s \boldsymbol{V}$ does not work in collinear superspace because $\bar{\D}^2 = 0$.
A new approach is required, which is the subject of \Sec{sec:Gaugetheory}.

A key ingredient for understanding gauge theories is a new type of superfield which does not have a counterpart in ordinary superspace.
This is a representation that is simultaneously chiral, anti-chiral, and real:
\begin{align}
\bar{\D}\s \RCA = 0, \qquad \D\s \RCA^\dagger = 0, \qquad \RCA^\dagger = \RCA\,,
\end{align}
where the symbol $\RCA$ was chosen since this will encode residual gauge transformations in light-cone gauge.
The chirality condition implies that $\RCA$ can be written as 
\begin{align}
\RCA\big(x,\eta,\eta^\dagger\big) &= \omega(x) + i\s \eta\, \xi\, \psi_\omega(x) + i\s \eta^\dagger \eta \, \PP \omega(x)\,,
\end{align}
for the bosonic scalar field $\omega$ and the fermionic scalar field $\psi_\omega$.
The reality condition implies
\begin{align}
\label{eq:RCA}
\omega = \omega^*, \qquad \psi_\omega = 0, \qquad \PP\s \omega = - \PP\s \omega^* = -\PP\s \omega = 0 \qquad \Longrightarrow \qquad \RCA\big(x,\eta,\eta^\dagger\big) = \omega(x)\,.
\end{align}
In the full $\mathcal{N} = 1$ superspace, this would just be trivial constant superfield.
In collinear superspace, $\PP_\perp \omega \neq 0$; this will turn out to be exactly the component we need to encode the superfield gauge transformations.

\subsection{Collinear Superspace Translations} 
\label{sec:SUSYTrans}

Under an ordinary SUSY transformation, the superspace coordinates transform as
\begin{align}
\theta^\alpha &\susy \theta^\alpha + \zeta^\alpha \,, \nonumber\\
\theta^{\dagger \dot{\alpha}} &\susy \theta^{\dagger \dot{\alpha}} + \bar{\zeta}^{\dot{\alpha}} \, , \nonumber\\
x^\mu &\susy x^\mu +  i\s\zeta\sigma^\mu\theta^{\dagger}+i\s\bar{\zeta}\bar{\sigma}^\mu \theta \, ,
\end{align}
where $\zeta^\alpha$ is a constant two-component Grassmann spinor.
To capture the same information in collinear superspace, we simply make the replacement $\theta^\alpha = \xi^\alpha\, \eta$ and $\zeta^\alpha = \xi^\alpha\, \epsilon$, which gives a representation of the collinear SUSY algebra in \Eq{eq:reducedSUSYalg}:  
\begin{align}
\eta &\susy \eta + \epsilon \, , \nonumber \\ 
\eta^\dagger &\susy \eta^\dagger + \epsilon^\dagger \,, \nonumber \\ 
x^\mu &\susy x^\mu + i \s\bar{n}^\mu \big(\epsilon\, \eta^\dagger  + \epsilon^\dagger\, \eta\big)\,.
\end{align}
We emphasize that $\epsilon$, which parametrizes collinear SUSY transformations, does not carry a spinor index.

Acting on a chiral superfield from \Eq{eq:chiralsuperfield}, a collinear SUSY transformation yields
\begin{align}
\label{eq:susyvariation}
\delta_\epsilon\s \bPhi & = - i\s \big(\epsilon\, \Q + \epsilon^\dagger\, \Q^\dagger \big) \bPhi \nonumber \\[5pt]
 & = \sqrt{2} \, \epsilon \, u + 2 \s i \, \epsilon^\dagger\, \eta\,  \PP\s \phi + \sqrt{2}\s i \, \epsilon \, \eta^\dagger \eta \, \PP \s u\,,
\end{align}
from which we can deduce the component transformations, 
\begin{align}
\delta_\epsilon \phi &= \sqrt{2} \, \epsilon \, u\,, \nonumber \\
\delta_\epsilon u &= - i\s\sqrt{2} \, \epsilon^\dagger\, \PP \s\phi\,,
\label{eq:componentshift}
\end{align}
with similar results for the conjugate fields.
As for ordinary chiral multiplets, we can introduce a shifted spacetime coordinate to simplify SUSY manipulations:
\begin{equation}
y^\mu \equiv x^\mu + i \s\bar{n}^\mu\, \eta^\dagger \eta\,,  \qquad y^\mu \susy y^\mu + 2\s i\, \bar{n}^\mu\, \epsilon^\dagger\, \eta\,.
\end{equation}
From this, it is clear that
\begin{equation}
\boldsymbol{\Phi}\big(x,\eta,\eta^\dagger\big) =\boldsymbol{\Phi}\big(y,\eta\big) = \phi(y) + \sqrt{2} \, \eta \, u(y)\,,
\end{equation}
which gives a slightly simpler way to derive \Eq{eq:componentshift}.  That said, we will stick with the $x^\mu$ coordinates throughout this paper.

Note that the highest component of a collinear chiral superfield -- the fermionic $u$ component -- transforms as a total derivative.
Because it is fermionic, though, we cannot construct a collinear-SUSY-invariant action using a standard bosonic chiral superpotential.
In the companion paper \cite{Cohen:2019gsc}, we show how to construct a novel fermionic chiral superpotential, using fermionic chiral superfields whose highest component is bosonic.%
As shown in \Eq{eq:DPhicomponents}, this kind of object is what one gets from $\bar{\D} \s\bPhi^\dagger$.

Starting from a real collinear superfield from \Eq{eq:vector}, we can derive the component transformation rules:  
\begin{align}
\delta_\epsilon\s a &= i\! \left( \epsilon \, b - \epsilon^\dagger b^\dagger  \right)\,,  \nonumber \\
\delta_\epsilon\s b &=  - i\s \epsilon \, v + i\s \epsilon\, \PP\s a\, ,\nonumber \\
\delta_\epsilon\s v &=\epsilon\, \PP\s b + \epsilon^\dagger\, \PP\s b^\dagger\, .
\label{eq:componentshift_vector}
\end{align}
Here $b = \xi^\alpha b_\alpha$ and the highest component $v= v_\mu\,\xi\s \sigma^\mu\s \xi^\dagger$ is bosonic, real, and transforms as a total derivative, and we will use that to construct Lagrangians in \Secs{sec:BuldingL}{sec:Gaugetheory}.
For the superfield in \Eq{eq:RCA}, which is simultaneously chiral, anti-chiral, and real, it transforms as
\begin{align}
\delta_\epsilon \RCA = 2\s i\, \eta^\dagger\, \eta\, \epsilon^\dagger \,\xi^\dagger\, \PP \omega = 0\,,
\end{align}
implying that $\RCA$ is a supersymmetric object, though we have not found a way to construct nontrivial Lagrangians with it.

Finally for completeness, we can test whether the collinear SUSY sub-algebra closes, by showing that the commutator of two transformations is a spacetime translation along the light-cone direction.
Given two SUSY transformations $\delta_{\epsilon_1}$ and $\delta_{\epsilon_2}$ acting on the components of a chiral superfield, we find
\begin{align}
\Big[ \delta_{\epsilon_1}, \delta_{\epsilon_2} \Big] \phi &= 2  \left( \epsilon_2\, \epsilon^{\dagger}_{1} -  \epsilon_1\, \epsilon^{\dagger}_{2}\right) \PP u\,, \nonumber\\
\Big[ \delta_{\epsilon_1}, \delta_{\epsilon_2} \Big] u & =  2  \left( \epsilon_2\, \epsilon^{\dagger}_{1} -  \epsilon_1\, \epsilon^{\dagger}_{2}\right) \PP \phi \,,
\end{align}
as expected from \Eq{eq:reducedSUSYalg}.

\section{A Collinear Superspace Sextant: Reparametrization Invariance}
\label{sec:RPI}

The spinor projections in \Sec{subsec:spinor_proj} naively appear to be an explicit breaking of Lorentz symmetry, since they identify a preferred light-cone direction.
However, this breaking is artificial:  the choice of $\xi^\alpha$ and $\tilde{\xi}^\alpha$ is arbitrary since any light-cone choice would yield the same physics.
The redundancy of choosing a light-cone direction encodes the underlying Lorentz structure of the theory via the RPI transformations (see \App{app:RPIgen} for details).
For our purposes, RPI simply enforces that the physics must be unchanged by the choice of light-cone direction and therefore that every object decomposed in light-cone coordinates must have well-defined RPI transformation properties. 

\subsection{RPI Transformations of the Light Cone}

To derive the action of the RPI transformations, we need to identify transformations on $\xi^\alpha$ and $\tilde{\xi}^\alpha$ which preserve 
\begin{align}
\xi^\alpha \s\tilde{\xi}_\alpha = 1,
\label{eq:RPIcondition}
\end{align}
which is equivalent to $n\cdot \bar{n} = 2$ in \Eq{eq:chiXiCondition}.
The most general linear transformation on $\xi^\alpha$ and $\tilde{\xi}^\alpha$ is  
\begin{align}
\xi_\alpha \,\,&\longrightarrow\,\, a \, \tilde{\xi}_\alpha  +b \, \xi_\alpha \,,\label{eq:PreRPI-II}\\[5pt]
\tilde{\xi}_\alpha \,\,&\longrightarrow\,\, c \, \tilde{\xi}_\alpha+d\, \xi_\alpha\,, \label{eq:PreRPI-I}
\end{align}
where $a$, $b$, $c$, and $d$ are complex coefficients.
Maintaining \Eq{eq:RPIcondition} requires 
\begin{equation}
 a\s d-b\s c = 1.
\end{equation}
This implies that the group of transformations that maintain the normalization of the spinors are complex linear transformation with unit determinant, namely SL(2,$\mathbb{C}$).\footnote{More specifically, the group is projective because overall signs play no role.}
The six generators correspond to Lorentz transformations on the celestial sphere~\cite{Larkoski:2014bxa}, whose properties are reviewed in more detail in \App{app:RPIgen}.

These six transformations are usually grouped into three categories: 
\begin{align}
\xi & \RPIi  \xi \,, & \tilde{\xi} &\RPIi \tilde{\xi}+ \kappa_{\rm I} \,  \xi\,, \label{eq:RPI1}\\[5pt]
\quad \quad \xi & \RPIii  \xi+ \kappa_{\rm II} \, \tilde{\xi}\,,& \tilde{\xi} &\RPIii  \tilde{\xi}\,, \label{eq:RPI2}\\[5pt]
\quad \quad \xi &\RPIiii   e^{-\kappa_{\rm III}/2} \, \xi\,, &\tilde{\xi} &\RPIiii  e^{\,\kappa_{\rm III}/2} \, \tilde{\xi}\,. \label{eq:RPI3}
\end{align}
While $\kappa_{\rm I}$ and $\kappa_{\rm II}$ are in general complex, we typically restrict $\kappa_{\rm III}$ to be real, since a simple phase rotation of $\xi$ and $\tilde{\xi}$ does not change $n^\mu$ or $\bar{n}^\mu$, as is clear from \Eq{eq:twospinorsn}.
One can also understand the reality of $\kappa_{\rm III}$  by examining the algebra given in \App{app:RPIgen}, or by recognizing that imaginary $\kappa_{\rm III}$ corresponds to the SO(2) little group.
Thus, there are five non-trivial RPI generators, which correspond to three different ways of maintaining \Eq{eq:RPIcondition}.
Taking $\xi$ to be fixed while shifting $\tilde{\xi}$ in the perpendicular direction yields RPI-I.
Reversing the roles of $\xi$ and $\tilde{\xi}$ yields RPI-II.
If both spinors transform by equal and opposite scale transformations, this yields RPI-III.
%

\subsection{RPI Transformations of Projected Objects}
To derive the action of RPI on projected objects, we simply apply the transformations for $\xi$ and $\tilde{\xi}$, while leaving the underlying Lorentz-covariant objects unchanged.
The relevant RPI transformations of various objects are summarized in \Tab{table:RPI-components}.

\begin{table}[t!]
\renewcommand{\arraystretch}{1.7}
\setlength{\arrayrulewidth}{.3mm}
\centering
\setlength{\tabcolsep}{0.95 em}
\begin{tabular}{ |c || c | c | c|}
\hline
 Object &  RPI-I &  RPI-II & RPI-III   \\ \hline \hline
 $\tilde{\xi}^\alpha $  & $\tilde{\xi}^\alpha + \rpii \,\xi^\alpha$   & $\tilde{\xi}^\alpha $ &  $ e^{\rpiiii/2}\, \tilde{\xi}^\alpha $ \\
$\xi^\alpha  $ &     $\xi^\alpha$    & $\xi^\alpha+ \kappa_{\rm II} \, \tilde{\xi}^\alpha$ &  $ e^{-\rpiiii/2}\, \xi^\alpha$  \\
\hline
  $\tilde{\PP} $ &  $ \tilde{\PP} +\rpii\, \PP_\perp+\kappa_{\rm I}^*\, \PP^*_\perp$  & $\tilde{\PP}$ & $ e^\rpiiii \,\tilde{\PP}$ \\
  $\PP  $ &  $\PP$   & $\PP + \kappa_{\rm II}^* \, \PP_\perp + \kappa_{\rm II} \, \PP_\perp^*$  &$e^{-\rpiiii}\,\PP$ \\
 $\PP_\perp $ &   $ \PP_\perp + \kappa_{\rm I}^*\, \PP$   &$\PP_\perp + \kappa_{\rm II} \, \tilde{\PP} $& $ \PP_\perp $ \\
  $\PP_\perp^* $ &  $\PP_\perp^* + \rpii\, \PP$  & $\PP_\perp^* + \kappa_{\rm II}^*  \, \tilde{\PP}$& $ \PP_\perp^* $ \\
\hline
$\phi  $ &    $\phi$   &  $\phi$ &  $ \phi$  \\
  $ u $ &   $u$ & $u + \kappa_{\rm II} \, \tilde{u}$ & $  e^{-\rpiiii/2}\,u $ \\
  $ \tilde{u} $ &     $ \tilde{u} + \kappa_{\rm I} \, u$ & $\tilde{u}$ & $  e^{\rpiiii/2}\, \tilde{u} $ \\
  \hline
  ${n} \cdot A $ &  $ {n} \cdot A +\sqrt{2} \left(\rpii\, \alc +\kappa_{\rm I}^*\, \alc^*\right)$  & ${n} \cdot A$ & $ e^\rpiiii \,{n} \cdot A$ \\
  $\bar{n} \cdot A   $ &  $\bar{n} \cdot A$   & $\bar{n} \cdot A + \sqrt{2} \left(\kappa_{\rm II}^* \, \alc + \kappa_{\rm II} \, \alc^* \right)$  &$e^{-\rpiiii}\,\bar{n} \cdot A $ \\
  $\alc  $ &  $\alc+\frac{\kappa_{\rm I}^*}{\sqrt{2}} \,\bar n \cdot A$     &  $\alc + \frac{\kappa_{\rm II}}{\sqrt{2}} \, n\cdot A$ &  $\alc$  \\
$\alc^*  $ &  $\alc^*+\frac{\rpii}{\sqrt{2}}\, \bar n\cdot A$  & $\alc^* + \frac{\kappa_{\rm II}^*}{\sqrt{2}} \,n\cdot A $ & $\alc^*$  \\
\hline
\end{tabular}
\caption{\small{The RPI transformation properties for various spinor projections, derivatives, and component fields.  RPI-II transformations can be derived for the component fields, but not for collinear superfields. }
\label{table:RPI-components}}
\end{table}

Under RPI-I and RPI-II, the light-cone four-vectors transform as 
\begin{align}
n^\mu &\RPIi n^\mu + \Delta_\perp^\mu \,, &  \bar{n}^\mu &\RPIi \bar{n}^\mu\,,\\[5pt]
 n^\mu &\RPIii n^\mu\,, & \bar{n}^\mu &\RPIii \bar{n}^\mu  + \epsilon^\mu_\perp\,, \label{eq:RPI2_nn}
\end{align}
where we have defined
\begin{align}
\label{eq:deltaepsilon}
\Delta^\mu_\perp = \kappa_{\rm I} \, \xi\s \sigma^\mu\s \tilde{\xi}^\dagger + \kappa_{\rm I}^* \, \tilde{\xi} \s \sigma^\mu\s \xi^\dagger \quad \text{and} \quad \epsilon^\mu_\perp = \kappa_{\rm II} \,  \xi \s\sigma^\mu \s\tilde{\xi}^\dagger + \kappa_{\rm II}^* \, \tilde{\xi}\s \sigma^\mu\s \xi^\dagger. 
\end{align}
The four-vectors $\Delta_\perp^\mu$ and $\epsilon_\perp^\mu$ only have non-zero components in the directions perpendicular to the light-cone, so RPI-I and RPI-II correspond to rotations around the light-cone~\cite{Marcantonini:2008qn}.\footnote{To make contact with the notation used in the SCET literature, simply replace $\Delta_\perp \cdot \partial =   \kappa_{\rm I}\, \PP_\perp + \kappa_{\rm I}^* \,\PP_\perp^* $ and  $\epsilon_\perp \cdot \partial =  \kappa_{\rm II}\, \PP_\perp^* + \kappa_{\rm II}^*\, \PP_\perp $, as can be verified using \Eqs{eq:Funnyd}{eq:deltaepsilon}.  This helps confirm that the generators we identify as RPI-I, -II, and -III correspond to the usual ones in the SCET literature.}
Under RPI-I and RPI-II, 
\begin{align}
\PP &\RPIi \PP \,,  & \tilde{\PP} &\RPIi \tilde{\PP} + \rpii\, \PP_\perp + \rpii^*\, \PP^*_\perp\, , \\
\PP &\RPIii\PP + \kappa_{\rm II}^*\, \PP_\perp + \kappa_{\rm II} \,\PP_\perp^* \,, & \tilde{\PP} &\RPIii \tilde{\PP} \,.
\end{align}
The mixed spinor derivatives transform as
\begin{align}
\PP_\perp &\RPIi \PP_\perp + \kappa_{\rm I}^*\, \PP \,,  & \PP_\perp^* & \RPIi \PP_\perp^*  + \kappa_{\rm I}\, \PP \,, \\
 \PP_\perp &\RPIii \PP_\perp  + \kappa_{\rm II}\, \tilde{\PP} \, , & \PP_\perp^* &\RPIii \PP_\perp^* +  \kappa_{\rm II}^*\, \tilde{\PP} \,.
\end{align} 

We can repeat the above logic for RPI-III, yielding
\begin{align}
n^\mu = \tilde{\xi}\s \sigma^\mu\s \tilde{\xi}^\dagger &\RPIiii e^{\,\kappa_{\rm III}/2}\, \tilde{\xi}\s \sigma^\mu\, e^{\,\kappa_{\rm III}/2}\, \tilde{\xi}^\dagger = e^{\,\kappa_{\rm III}}\, n^\mu,\\[5pt]
\bar{n}^\mu = \xi\s \sigma^\mu\s \xi^\dagger &\RPIiii  e^{- \kappa_{\rm III}/2}\,\xi\s \sigma^\mu \, e^{- \kappa_{\rm III}/2} \xi^\dagger = e^{- \kappa_{\rm III}}\, \bar{n}^\mu,
\end{align}
which correspond to boosts along the light-cone direction~\cite{Marcantonini:2008qn}.
Therefore, $\PP =  \bar{n} \cdot \partial$ and $\tilde{\PP} =  n \cdot \partial$ defined in \Eq{eq:Funnyd} transform as
\begin{equation}
\PP \RPIiii e^{- \kappa_{\rm III}}\, \PP\,, \qquad\text{and}\qquad \tilde{\PP} \RPIiii e^{\,\kappa_{\rm III}}\, \PP\,,  
\end{equation}
since $\partial_\mu$ is an ordinary Lorentz vector that is unaffected by RPI.
Note that $\PP_\perp$ and $\PP_\perp^*$ are invariant under RPI-III, as they contain both $\xi^\alpha$ and $\tilde{\xi}^\alpha$.
Additionally, we see that $\Box = \PP\s \tilde{\PP}  - \PP_\perp^*\s \PP_\perp$ is invariant under all of RPI-I, RPI-II, and RPI-III, which is an important consistency check.

\subsection{RPI Transformations of Component Fields}
\label{subsec:RPIcomponents}

The RPI transformation properties of component fields can also be derived from the transformations on $\xi$ and $\tilde{\xi}$.
A scalar field $\phi$ transforms trivially.
The fermion field $u = \xi^\alpha u_\alpha$ is invariant under RPI-I
\begin{align}
u \RPIi u\,,
\end{align}
as can be derived using \Eq{eq:RPI1}, and transforms as 
\begin{align}
u \RPIiii e^{-\kappa_{\rm III}/2}\, u\,,
\end{align} 
as is clear from \Eq{eq:RPI3}.
As will be discussed below in \Sec{subsec:RPIii}, $u$ transforms into $\tilde{u} \equiv  \tilde{\xi}^\alpha \s u_\alpha$ under RPI-II: 
\begin{equation}
\label{eq:RPI2_on_u}
u \RPIii u + \kappa_{\rm II} \,\tilde{u}\,.
\end{equation}
Note that the $\tilde{u}$ field does not appear explicitly in the constructions in this paper.
This is connected to the fact that we will be forced to check RPI-II directly on the component Lagrangian, as explained below in \Sec{subsec:RPIii}.

As discussed in \Sec{sec:ProjectGauge}, the propagating modes of a gauge field are naturally expressed as a complex scalar 
\begin{equation}
\label{eq:scriptAdef}
\mathcal{A} \equiv A_\mu \, \xi\s \sigma^\mu\s \tilde{\xi}^\dagger\,.
\end{equation}
However, since this ``scalar'' explicitly depends on $\xi$ and $\tilde{\xi}$, it has non-trivial RPI transformations.  
In particular, $\mathcal{A}$ is not invariant under RPI-I or RPI-II, although it is invariant under RPI-III:
\begin{align}
\mathcal{A} &\RPIi \mathcal{A} + \frac{\kappa_{\rm I}}{\sqrt{2}} \, \bar{n}\cdot A\,,\\[2pt]
\mathcal{A} &\RPIii \mathcal{A} + \frac{\kappa_{\rm II}}{\sqrt{2}} \, n\cdot A\,,\\[2pt]
\mathcal{A} &\RPIiii \mathcal{A} \,.
\end{align}
In light-cone gauge where $\bar{n}\cdot A = 0$, $\mathcal{A}$ is invariant, which is useful for writing a gauge theory Lagrangian that is consistent with RPI-I transformations.
This observation will allow us to construct theories in collinear superspace that preserve both RPI-I and RPI-III, while again RPI-II must be checked at the component level.

\subsection{Implications for Collinear Superspace}
\label{subsec:ImpCollSup}

\begin{table}[t!]
\renewcommand{\arraystretch}{1.8}
\setlength{\arrayrulewidth}{.3mm}
\centering
\setlength{\tabcolsep}{0.95 em}
\begin{tabular}{ |c || c | c|}
    \hline
 Field &  RPI-I &   RPI-III   \\ \hline \hline
     $\eta $  & $\eta$   &    $ e^{\rpiiii/2}\, \eta $ \\
     \hline
      $\Q$ & $\Q$& $e^{-\kappa_{\rm III}/2}  \Q$ \\
     \hline
      $\D$ &$\D$ & $e^{-\kappa_{\rm III}/2} \,\D$ \\
  $\bar{\D}$ & $\bar{\D}$ & $e^{-\kappa_{\rm III}/2} \,\bar{\D}$ \\
  \hline
$\bPhi$  & $\bPhi$   &  $\bPhi$    \\
\hline
\end{tabular} 
\caption{\small{RPI transformations for various collinear superspace objects after setting $\tilde{\eta} = 0$.  Note we have not provided the RPI-II transformations since they take us outside the collinear SUSY sub-algebra. }
}
\label{table:RPI-superfields}
\end{table}

Due to the connection between SUSY and Lorentz invariance, clearly the superspace coordinates $\eta$ and $\tilde{\eta}$ must transform non-trivially under RPI.
Using their definitions in \Eq{eq:eta_def} and remembering that a Lorentz spinor -- specifically $\theta_\alpha$ for our purposes here -- is invariant under RPI, implies
\begin{align}
\eta &\RPIi \eta - \kappa_{\rm I} \,  \tilde{\eta} \,, &  \tilde{\eta} &\RPIi \tilde{\eta} \,, \\[2pt]
\eta &\RPIii \eta\,, &  \tilde{\eta} &\RPIii \tilde{\eta}- \kappa_{\rm II} \, \eta\,, \label{eq:RPI2_eta}\\[2pt]
\eta &\RPIiii e^{\,\kappa_{\rm III}/2}\, \eta\,,&  \tilde{\eta} &\RPIiii e^{-\kappa_{\rm III}/2} \, \tilde{\eta}  \,.
\end{align}
Additionally, the collinear SUSY generators $\Q$ and $\Q^\dagger$ have non-trivial transformation properties under RPI, as they must since the ordinary SUSY generators transform as Lorentz spinors.

We immediately see that setting $\tilde{\eta} = 0$ is compatible with RPI-I and RPI-III, but not with RPI-II.
The reason is that the shift required by \Eq{eq:RPI2_eta} generically makes $\tilde{\eta}$ non-zero.
Therefore, as discussed further in \Sec{subsec:RPIii} below, we cannot make RPI-II manifest in collinear superspace.

There is a reduced RPI compatible with collinear superspace, consisting of just RPI-I and RPI-III: 
\begin{align}
\Aboxed{\begin{minipage}{0.185\linewidth}\vspace{7pt}$\hspace{7pt}\text{taking}\quad\tilde{\eta} = 0$\vspace{7pt}\end{minipage}} \hspace{50pt}  \eta  \RPIi \eta, \qquad   \eta \RPIiii e^{\,\kappa_{\rm III}/2} \, \eta\,.
\end{align}
Note that RPI-III acts like an imaginary $R$-symmetry where $\eta$ has $R$-charge $+1/2$.
Related transformation properties are inherited by the collinear super-covariant derivatives from \Eq{eq:defDDBar}:
\begin{align}
\D \RPIiii e^{-\kappa_{\rm III}/2} \,\D\,, \qquad \text{and} \qquad \bar{\D} \RPIiii e^{-\kappa_{\rm III}/2}\, \bar{\D}\,.
\end{align}
The RPI-I and RPI-III transformation properties of various collinear superspace objects are given in \Tab{table:RPI-superfields}.

In order for RPI-I and RPI-III to be manifest at the Lagrangian level, collinear superfields have to have well-defined transformation properties.
Because the lowest component of a standard chiral multiplet is a Lorentz scalar, we expect $\boldsymbol{\Phi}$ from \Eq{eq:chiralsuperfield} to be invariant under both RPI-I and RPI-III.
This can be verified explicitly using the component transformation properties from \Tab{table:RPI-components}.
For example, performing an RPI-III transformation, we find
\begin{align}
\phi &\RPIiii  \phi \, , \nonumber\\[2pt]
\eta \, u &\RPIiii  e^{\,\kappa_{\rm III}/2}\, \eta\,  e^{-\kappa_{\rm III}/2}\, u = \eta \, u \, ,\nonumber\\[2pt]
\eta^\dagger \eta \, \PP \phi &\RPIiii  e^{\,\kappa_{\rm III}/2}\,\eta^\dagger\, e^{\,\kappa_{\rm III}/2} \, \eta \, e^{- \kappa_{\rm III}}\, \PP \phi = \eta^\dagger \eta \, \PP \phi\,.
\end{align}
Note that $\kappa_{\rm III}$ is real, so $\eta^\dagger\s \eta$ is not RPI-III invariant on its own.
A similar calculation shows that each term of $\bPhi$ is RPI-I invariant as well.
We emphasize that $\bPhi$ \emph{does not} have well-defined superfield RPI-II transformation properties, though its components do.

\subsection{Where is RPI-II?}
\label{subsec:RPIii}

Ultimately, we are interested in constructing Lorentz-invariant theories, so we want to enforce the full RPI symmetry, including RPI-II.
Because RPI-II and collinear SUSY do not commute, though, we cannot simultaneously realize RPI-II while imposing the defining constraint of collinear superspace:  $\tilde{\eta} = 0$.
The reason is that RPI-II corresponds to a translation of $\tilde{\eta}$, as can be seen in \Eq{eq:RPI2_eta}.
Therefore, unlike for RPI-I and RPI-III, there are no EFT rules for constructing RPI-II-invariant operators directly in collinear superspace.

Of course, what is really going on is that RPI-II and collinear SUSY are just non-commuting sub-algebras of a larger $\mathcal{N} = 1$ structure.
After all, collinear SUSY (two supercharges) plus full RPI implies at least $\mathcal{N} = 1$ SUSY (four supercharges), since that is the smallest graded algebra consistent with Lorentz invariance \cite{Haag:1974qh}.
So while RPI-II does not map collinear superfields to collinear superfields, we can still apply the RPI-II transformations from \Tab{table:RPI-components} to component fields.
We will later use these component transformations to show that RPI-II is respected by the Lagrangians constructed in \Secs{sec:BuldingL}{sec:Gaugetheory}.

From \Tab{table:RPI-components}, we see that $u$ transforms into $\uu$ under RPI-II.
The field $\uu$ has a ``top-down" interpretation as the helicity component (of a massless spinor representation of the Lorentz algebra), which is non-propagating when we construct a theory on the light cone.
From the ``bottom-up'' perspective, we can view $\uu$ as a constrained fermion mode in the effective theory, whose constraint equation must be the most general one allowed by the symmetries of the theory, namely RPI-I and RPI-III.
These two perspectives are of course related, where the bottom-up constraint should corresponds to the top-down equation of motion used to integrate out $\uu$.

Assuming $\uu$ is linear in $u$, we can derive a constraint equation of the form $\uu + \hat{\mathcal{O}}\s u = 0$, where $\hat{\mathcal{O}}$ is some differential operator.
To ensure RPI-III invariance, we need all terms in this constraint equation to have the same RPI-III charge, so that we may consistently set it to zero.
Since $u$ ($\uu$) has RPI-III charge $-1/2$ ($+1/2$), $\hat{\mathcal{O}}$ must have RPI-III charge $+1$, which means that $\hat{\mathcal{O}}$ must be proportional to $\tilde{\PP}$ or $1/\PP$.
Note that $\PP_\perp$ and $\PP_\perp^*$ have no RPI-III charge, so we can use them freely as long as $\hat{\mathcal{O}}$ has mass dimension zero.

Turning to RPI-I, $u$ is inert but $\uu \to \uu + \kappa_{\rm I} \s u$, so we must have $\hat{\mathcal{O}} \rightarrow \hat{\mathcal{O}} - \kappa_{\rm I}$.
Assuming there are no mass scales in the problem, this uniquely fixes $\hat{\mathcal{O}} = - \PP_\perp^* / \PP$, yielding the constraint $\uu = (\PP_\perp^*/\PP) u$.%
\footnote{More generally, $\hat{\mathcal{O}}$ could include a term proportional to $f/\PP$ if $f$ has mass dimension one.}
Inserting this into \Eq{eq:RPI2_on_u} yields the RPI-II transformation: 
\begin{align}
\label{eq:uNLRPIii}
u  &\RPIii u +  \kappa_{\rm II}\s \frac{\PP_\perp^*}{\PP}\s u\,.
\end{align}
One can verify that this transformation is consistent with the RPI algebra given in \Eq{eq:RPIComm}.

For the massless theories in \Secs{sec:BuldingL}{sec:Gaugetheory}, we can use \Eq{eq:uNLRPIii}  to verify the RPI-II invariance of our derived Lagrangians at the component level.
For theories that involve additional mass scales, the RPI-II transformations of $u$ cannot be uniquely defined using the above logic.
This is one of the reasons why in the companion paper, instead of imposing a constraint on $\uu$, we introduce a novel superfield whose lowest component is $\uu$ \cite{Cohen:2019gsc}.
Here, our focus is on massless theories, so we can simply use \Eq{eq:uNLRPIii} without reference to $\uu$.%
\footnote{While this  realization of RPI-II might seems odd, is not unexpected given that the opposite helicity field $\tilde{u}$ is absent from the theory. Indeed, the RPI-II transformations of an uncharged fermion in SCET can be derived by demanding that the full theory fermion be invariant \cite{Cohen:2016jzp}.}

\section{Learning the Ropes:  Free Chiral Multiplets}
\label{sec:BuldingL}

In the spirit of EFTs, our goal is to elucidate the underlying symmetries and power-counting rules that yield valid Lagrangians in collinear superspace.
We are now armed with all the necessary tools to understand what superspace operators are allowed without having to rely on matching to an explicitly Lorentz-invariant construction as in \Refs{Cohen:2016jzp, Cohen:2016dcl}.  
Using the ingredients from \Secs{sec:formalism}{sec:RPI}, we can construct Lagrangians directly in collinear superspace by demanding RPI-I, RPI-III, collinear SUSY, and global/gauge symmetries.
As emphasized in \Sec{subsec:RPIii}, imposing RPI-II requires explicit manipulation of the component Lagrangian.
In this section, we consider the simplest case of a single free chiral multiplet, which already illuminates many aspects of the collinear superspace formalism.

\subsection{Rules for Building an Action}

The most straightforward way to impose collinear SUSY on an action is to express the Lagrangian as the lowest component of a total superspace derivative:%
\footnote{An equivalent way to write \Eq{eq:genericL} is \begin{align}\mathcal{L} = \int \text{d} \eta \s \text{d} \eta^\dagger \, \bV_{\rm comp} + \int \text{d} \eta \, \bPhi_{\rm comp} + \int \text{d} \eta^\dagger \,\bPhi^\dagger_{\rm comp}\,,\end{align} though we found \Eq{eq:genericL} to be more convenient for practical calculations.}
\begin{align}
\label{eq:genericL}
\mathcal{L} =  \Big( \D \s\bar{\D}\s \bV_{\rm comp} + \D \s\bPhi_{\rm comp} + \Dbar \s\bPhi^\dagger_{\rm comp}  \Big) \Big|_{0}\,\,,
\end{align}
where $\bV_{\rm comp}$ is a composite real multiplet $\big(\bV_{\rm comp} = \bV^\dagger_{\rm comp}\big)$, $\bPhi_{\rm comp}$ is a composite chiral multiplet $\big(\Dbar\s \bPhi_{\rm comp} = 0\big)$, and the zero subscript indicates the restriction to $\eta = 0 = \eta^\dag$. 
Here, ``composite'' means that it is constructed from elementary superfields, \emph{e.g.} a product of elementary chiral multiplets is a composite chiral multiplet.
Using \Eqs{eq:componentshift}{eq:componentshift_vector}, it is clear that the lowest components of $\D\s \Dbar\s \bV_{\rm comp}$ and $\D \s \bPhi_{\rm comp}$ transform as total derivatives under collinear SUSY.
Therefore, the action $S = \int \text{d}^4 x \, \mathcal{L}$ is invariant under collinear SUSY.
An analogous logic is used to justify the SUSY invariance of the standard off-shell superspace formulation of $\mathcal{N} = 1$ SUSY, see \emph{e.g.}~\cite{Bertolini:2013via}.

In this paper, we work exclusively with elementary superfields that are bosonic.
If $\bPhi_{\rm comp}$ is bosonic, then $\D \bPhi_{\rm comp}$ is fermionic, so it is not useful for our purposes here to include such a term in the action, \Eq{eq:genericL}. 
Therefore, we set $\bPhi_{\rm comp} = 0$ for the remainder of this paper, yielding the generic collinear SUSY Lagrangian
\begin{align}
\label{eq:VrestrictedL}
\mathcal{L} =  \D \bar{\D} \bV_{\rm comp} \Big|_{0}\,\,,
\end{align}
which effectively means that there is no analog for the ``superpotential'' in this construction.\footnote{If $\bPsi_{\rm comp}$ is bosonic and chiral, then $\bPhi_{\rm comp} = \Dbar \bPsi^\dagger_{\rm comp}$ is fermionic and chiral.  Inserting this $\bPhi_{\rm comp}$ into \Eq{eq:genericL} is equivalent to setting $\bV_{\rm comp} = \bPsi_{\rm comp} + \bPsi_{\rm comp}^\dagger$, and not only does it not generate a new type of term, but it in fact yields a total derivative.}

The Lagrangian must satisfy the following requirements, which impose a set of constraints on the form of $\bV_{\rm comp}$:
\begin{itemize}
\item \textbf{Mass dimension four}:
Recall that $\big[\D\big] = \big[\bar{\D}\big] = 1/2$, which implies that $\bV_{\rm comp}$ must have mass dimension three.%
\footnote{In this paper, we perform power counting based on mass dimension, which one can show is equivalent to the SCET power counting $(\tilde{\PP}, \PP, \PP_\perp) \sim Q (\lambda^2, 1, \lambda)$ when RPI-III is taken into account.}
The kinetic term for chiral superfields is given in \Eq{eq:VcompKinetic} below, along with arguments for its validity and uniqueness.
\item \textbf{Lorentz invariant}:
Lorentz invariance of the $S$-matrix is equivalent to RPI as discussed in \Sec{sec:RPI}.
Since $\D$ and $\Dbar$ are invariant under RPI-I, $\bV_{\rm comp}$ must be as well.
The product $\D \s\Dbar$ has RPI-III charge $-1$, so $\bV_{\rm comp}$ must have RPI-III charge $+1$.
As discussed in \Sec{subsec:RPIii}, RPI-II has to be checked at the component level and cannot be directly enforced as a property of $\bV_{\rm comp}$.
\item \textbf{Gauge invariant}:
To have an RPI-I invariant action, we must fix to light-cone gauge, as discussed around \Eq{eq:scriptAdef}.
In \Sec{sec:Gaugetheory}, we show that a residual gauge symmetry survives in the form of a real, chiral, and anti-chiral multiplet $\RCA$.  Enforcing residual gauge symmetry yields additional constraints on the Lagrangian.
\end{itemize}
As in standard EFTs, we can write down a valid collinear SUSY action by identifying all terms consistent with these requirements.
Unlike standard EFTs (but familiar from SCET), the action will contain inverse momentum scales, which nevertheless yields a local $S$-matrix as it must since this construction is equivalent to the manifestly local off-shell superspace description.

\subsection{Constructing the Kinetic Term}
\label{subsec:chiral_kinetic}

We next want to build the kinetic term for the chiral superfield defined in \Eq{eq:chiralsuperfield}.
As we argue in the following, the unique kinetic term allowed by the criteria listed above is: 
\begin{equation}
\label{eq:VcompKinetic}
\bV_{\rm comp} =  \frac{i}{2}\bPhi^\dagger \frac{\Box}{\PP}  \bPhi\,.
\end{equation}
Despite the inverse momentum scale, this kinetic term is in fact local since $\Box$ is parametrically small compared to $\PP$, in keeping with the analysis of \Refs{Cohen:2016jzp, Cohen:2016dcl}.
Note that $\bV_{\rm comp}$ in \Eq{eq:VcompKinetic} is bosonic, real,%
\footnote{Strictly speaking, $\bV_{\rm comp}$ is only real up to total derivative terms appearing in the Lagrangian.}
and has mass dimension three, thereby following the rules outlined in the previous section.

Writing out the kinetic term in components, we have  
\begin{align}
\label{eq:SUSYSCETLagrangian}
\mathcal{L}  =\D \bar{\D} \bV_{\rm comp} \,\Big|_{0} 
& =\frac{i}{2}\,  \D \s\bar{\D} \bigg[ \bPhi^\dagger \frac{\Box}{\PP}  \bPhi \bigg] \bigg|_{0}   \notag\\[10pt]
& =  \bigg[- \,\bPhi^\dagger\s  \Box\s \bPhi +\frac{1}{2} \big( \bar{\D}\s  \bPhi^\dagger \big)\frac{i\s \Box}{\PP} \big(\D\s \bPhi \big)\bigg]\, \bigg|_{0}  \notag\\[10pt]
& = -\phi^* \Box \phi + i\s u^\dag \frac{\Box}{\PP} u\,,
\end{align}
which are the canonical  light-cone kinetic terms for the $\phi$ and $u$ fields.
In the second line, we made use of the product rule $\D\s(\mathbf{X\s Y}) = (\D\s \mathbf{X})\mathbf{Y} \pm \mathbf{X}(\D\s \mathbf{Y})$, where the sign depends on whether $\mathbf{X}$ is bosonic $(+)$ or fermionic $(-)$.
We also used the anti-commutation relation and chiral condition to write $\D \s\Dbar\s \bPhi^\dagger = -2\s i\, \PP\s \bPhi^\dagger$.

It is illustrative to explain in detail why \Eq{eq:VcompKinetic} is the unique kinetic term, since similar arguments will be applied in \Sec{sec:Gaugetheory}.
The kinetic term has to be bilinear in superfields, with one chiral and one anti-chiral field to make sure that $\D\s \bar{\D}\s \bV_{\rm comp}$ does not vanish.
Naively, the closest analog to the usual canonical K\"ahler potential would be $\bV_{\rm comp} = \bPhi^\dagger\s \bPhi$, but this is disqualified since it has mass dimension two instead of three.
Additionally, this term has the wrong RPI-III transformation since $\bV_{\rm comp} \to e^{-\rpiiii} \,\bV_{\rm comp}$ is required to balance $\D\s\bar{\D} \to e^\rpiiii \,\D\s\bar{\D}$ to obtain an invariant action, whereas $\bPhi^\dagger \s\bPhi$ is invariant.
We can compensate for this by including either $\tilde{\PP}$ or $1/\PP$, both of which have RPI-III charge $+1$, but only $1/\PP$ is invariant under RPI-I.\footnote{It is interesting that the action is forced to have inverse momentum scales by RPI.  For example, the local operator $\bV_{\rm comp} = \bPhi\, \Dbar\s \D\,  \bPhi^\dagger = -2\s i\, \bPhi\, \PP\s\bPhi^\dagger$ has mass dimension three but has RPI-III charge $-1$ instead of $+1$.}
Then to achieve the correct mass dimension, we can insert factors of the RPI invariant $\Box$.
Altogether, this yields $\bV_{\rm comp} = \frac{i}{2}\s\bPhi^\dagger\s \frac{\Box}{\PP} \s\bPhi$ as claimed, with the factor of $i$ needed to ensure that $\bV_{\rm comp}^\dagger = \bV_{\rm comp}$ and the $1/2$ for canonical normalization of the kinetic terms.

One might be concerned that starting from the required bilinear $\bPhi^\dagger\s \bPhi$, there could be additional independent terms one could write down using alternate derivative choices.
However, the space-time derivatives $\tilde{\PP}$, $\PP_\perp$, and $\PP_\perp^*$ have non-trivial RPI-I transformations, and are as such not useful for constructing the kinetic term, since $\bPhi$ is RPI-I invariant.
Said another way, while it is possible for $\bV_{\rm comp}$ to involve $\tilde{\PP}$, $\PP_\perp$, or $\PP_\perp^*$ directly, integration by parts can always be used to combine them into $\Box$.
Regarding the super-covariant derivatives, note that since they are fermionic and $\bPhi$ is bosonic, $\D$ and $\bar{\D}$ have to come in pairs.  Then we can use $\{ \D, \bar{\D}  \} = -2\s i\,\PP$, integration by parts, and the chirality conditions to convert them to space-time derivatives.
Note that the discussion in this paragraph only holds for bilinear terms, where integration by parts is particularly powerful, but it will not hold in general, see~\Sec{subsect:KahlerDiscussion}.
%

\subsection{Verifying RPI-II}
\label{sec:revisitRPI2}

To verify that \Eq{eq:SUSYSCETLagrangian} satisfies RPI-II, we have to work directly in components.
Following \Sec{subsec:RPIii}, we assume that \Eq{eq:uNLRPIii} is the correct RPI-II transformation law.
The scalar kinetic term is manifestly RPI-II invariant.
We can check the fermion kinetic term by direct computation:
\begin{align}
\mathcal{L} \supset i \s u^\dag\s \frac{\Box}{\PP}\s u \RPIii & i \left( u^\dagger + \kappa^*_{\rm II} \frac{\PP_\perp}{\PP}\s   u \right) \left( \frac{\Box}{\PP} - \frac{\Box (\kappa_{\rm II}^*\, \PP_\perp + \kappa_{\rm II}\, \PP_\perp^*)}{\PP^2} \right) \left( u + \kappa_{\rm II}\, \frac{\PP_\perp^*}{\PP}\s u \right) \nonumber \\[8pt] 
&\quad= i \s u^\dagger\s \frac{\Box}{\PP}\s u + \mathcal{O}\big(\kappa_{\rm II}^2\big)\,,
 \label{eq:fermion_check_rpi_ii}
\end{align}
confirming that the full kinetic term for a collinear chiral superfield satisfies RPI-II.
Together with RPI-I, RPI-III, and collinear SUSY, this confirms that \Eq{eq:SUSYSCETLagrangian} describes a theory with full $\mathcal{N} = 1$ SUSY, albeit written in a language where Lorentz invariance and half of SUSY is obscured.

\subsection{Where is the K\"ahler Potential?}
\label{subsect:KahlerDiscussion}

We argued above that these constructions lack a ``superpotential,'' which means that we will need the new technology to be introduced in \Ref{Cohen:2019gsc} to write down mass terms and Yukawa interactions.%
\footnote{The Wess-Zumino model was studied in \Ref{Cohen:2016dcl}, but only with the help of external currents.}
In a similar spirit, it is natural to wonder if it is possible to write down a non-trivial ``K\"ahler potential,'' which would allow us to investigate higher-order interactions.

As a warm up, consider making the replacement $\Box \to \Box + m^2$ in \Eq{eq:VcompKinetic}.
This yields the Lagrangian
\begin{align}
\label{eq:SUSYSCETLagrangian_with_mass}
\mathcal{L}  = \D\s \bar{\D}\s \bV_{\rm comp} \Big|_{0}  \stackrel{?}{\supset} \frac{i}{2}\s \D\s \bar{\D}\s \bigg[ \bPhi^\dagger\s \frac{(\Box+m^2)}{\PP}  \bPhi \bigg] \bigg|_{0} = -\phi^* \big(\Box + m^2\big) \phi + i\s u^\dag \frac{\Box+m^2}{\PP} u,
\end{align}
which naively looks like a theory with mass terms.
However, since the fermion is now in a massive representation of the Lorentz group, its corresponding RPI transformations are not given by \Eq{eq:uNLRPIii}.
Specifically, the RPI-II transformations of $u$ must now depend on $m$, and this spoils the RPI invariance of the conjectured Lagrangian given in \Eq{eq:SUSYSCETLagrangian_with_mass}.
This should not come as a surprise, since from the top-down perspective, a mass term in SUSY yields non-trivial $F$-term equations of motion, which are absent from the present construction.

In fact, most non-canonical choices of $\bV_{\rm comp}$ will violate RPI-II in some way.
As a concrete example, consider a massless theory (such that \Eq{eq:uNLRPIii} still holds) with the following class of higher-dimension operators
\begin{equation}
\bV_{\rm comp} \stackrel{?}{\supset} \frac{i}{\Lambda^{(n+m-2)}} \big(\bPhi^\dagger\big)^n \s\frac{ \Box}{\PP}\s \big(\bPhi\big)^m + \text{h.c.} \, ,
\end{equation}
where $n$ and $m$ are integers and $\Lambda$ has mass dimension 1.
This term is bosonic and real, has mass dimension 3 and RPI-III charge $+1$, and is RPI-I invariant: it is therefore a candidate for inclusion in $\bV_{\rm comp}$.
By explicit computation, though, one can check that it violates RPI-II.
In fact, apart from introducing additional factors of $(\Box/\Lambda^2)$ into \Eq{eq:VcompKinetic}, we have been unable to identify any non-canonical $\bV_{\rm comp}$ that preserves RPI-II while still respecting RPI-I, RPI-III, and collinear SUSY.

From the bottom-up perspective, this simply emphasizes the importance of RPI-II in enforcing Lorentz invariance.
From the top-down perspective, it underscores an interesting fact about K\"ahler potentials.
Even in theories with a vanishing superpotential, non-canonical K\"ahler potentials generate non-zero $F$-terms proportional to fermion bilinears:
\begin{equation}
F^i = \frac{1}{2}\, \Gamma^i_{jk}\, \chi^j\, \chi^k,
\end{equation}
where $\Gamma^i_{jk}$ is the Christoffel connection derived from the K\"ahler metric.
Since our constructions lack auxiliary fields, we cannot generate such a term (at least not with a linear realization of collinear SUSY).

In the companion paper, we introduce superfields whose lowest component is $\tilde{u}$ and whose highest component is $F$, making it possible to realize non-trivial K\"ahler potentials (and superpotentials) after imposing RPI-II.
For this paper, though, we have only provided the technology for writing the Lagrangian for massless free chiral multiplets.
To obtain non-trivial interactions, we have to turn to gauge theories.

\section{Maiden Voyage:  Gauge Theories}
\label{sec:Gaugetheory}

Now that we have gained experience applying the EFT rules of collinear superspace to a free chiral multiplet, it is straightforward to explore the structure of gauge theories.
In this section, we explain how gauge invariance constrains operators in collinear superspace, starting from the simplest case of an Abelian gauge theory and then lifting to a non-Abelian gauge theory.
Obviously, the latter case requires introducing interactions, which provides a non-trivial check of the collinear superspace formalism.

As discussed in \Sec{sec:Superfields}, the familiar $\mathcal{N} =1$ method of organizing gauge degrees of freedom into a vector multiplet is not possible in collinear superspace.
That said, the physical polarizations of the gauge field and the gaugino can be packaged into a chiral superfield $\bPhiA$ whose kinetic term is given by \Eq{eq:SUSYSCETLagrangian}. 
In what follows, we demonstrate how a residual gauge symmetry, along with RPI, can be used to derive the rest of the gauge theory Lagrangian.

\subsection{Abelian Gauge Theory}
We begin with the Abelian case.  
As discussed in \Sec{sec:ProjectGauge}, our construction is based on a complex light-cone scalar field $\alc$ that is built from the two propagating gauge degrees of freedom.
Under RPI-I, $\alc$ has non-trivial transformation properties, so in order to package $\alc$ into a superfield, it is necessary to enforce light-cone gauge in collinear superspace, where $\bar{n}\cdot A = 0$ and $n\cdot A$ is non-propagating and therefore integrated out.
In light-cone gauge, $\alc$ is inert under both RPI-I and RPI-III.

Since we have written the gauge modes suggestively as a complex scalar $\alc$, it is clear how to package it into a gauge chiral superfield (see~\emph{e.g.}~\cite{Leibbrandt:1987qv} for a review):
\begin{align}
\label{eq:PhiA}
\bPhiA =\alc^* - \sqrt{2} i \eta \lambda^\dagger + i \eta^\dagger \eta \, \PP \alc^* \qquad &\text{with} \qquad \bar{\D} \s\bPhiA  = 0 \, , \notag\\[6pt]
\bPhiA^\dagger = \alc - \sqrt{2} i \eta^\dagger \lambda - i \eta^\dagger \eta \, \PP \alc  \hspace{32pt} &\text{with} \qquad \D\s \bPhiA^\dagger  = 0\,.
\end{align}
In analogy to \Eq{eq:udecom_reverse}, we have defined the propagating gaugino as $\lambda \equiv \xi^\alpha \lambda_\alpha$, which is operationally an anti-commuting scalar.
Note that in \Eq{eq:PhiA}, the chiral superfield contains the conjugate fields $\alc^*$ and $\lambda^\dagger$, and vice verse for the anti-chiral field.
This unusual organization of the degrees of freedom arises because one has to add $+1/2$ units of helicity to go from the lowest to highest component of a chiral multiplet, $0 \to +1/2$ for \Eq{eq:chiralsuperfield} and $-1 \to -1/2$ for \Eq{eq:PhiA}.
One can also understand this by matching to the full $\mathcal{N} = 1$ theory (see \Refs{Cohen:2016jzp, Cohen:2016dcl} and further discussion in \Ref{Cohen:2019gsc}).

Even after enforcing light-cone gauge, there is a residual gauge transformation on the chiral gauge superfield.
As mentioned in \Sec{sec:Superfields}, this can be parametrized by $\RCA$, a superfield that is both chiral and real (and therefore anti-chiral): 
\begin{align}
&\bPhiA \gauge \bPhiA + \s\PP_\perp^*\s \RCA \,, 
\label{eq:residualgaugetrans}
\end{align}
In components this yields,
\begin{align}
&\alc\,\, \gauge \alc +  \s \PP_\perp\s \omega \, , \hspace{27pt} \quad \lambda_\alpha\gauge \lambda_\alpha \,,
\end{align}
Note that the gaugino $\lambda$ does not transform since this is an Abelian model.
The transformation of the gauge scalar $\alc$ is inferred by inserting the standard gauge transformation $A_\mu \to A_\mu +  \s \partial_\mu \omega$ into \Eq{eq:def_script_A}.
Crucially, the residual gauge transformation $\omega$ is consistent with light-cone gauge.
To see this, note that 
\begin{align}
\label{eq:fullloretnzgaugetrans}
 \bar{n} \cdot A \gauge  \bar{n} \cdot A +  \PP \s \omega = \bar{n} \cdot A,
\end{align}
where in the last step we used the fact that $\omega$ is the lowest component of $\RCA$ and therefore satisfies $\PP \omega = 0$.
Thus, the light-cone gauge condition $\bar{n} \cdot A = 0$ is maintained by the residual gauge transformations defined in \Eq{eq:residualgaugetrans}.

Plugging the gauge chiral superfield into the chiral kinetic term from \Eq{eq:SUSYSCETLagrangian}, the component Lagrangian takes the desired form:
\begin{align}
\mathcal{L}= \frac{i}{2} \D\s \bar{\D} \bigg[ \bPhiA^\dagger \frac{\Box}{\PP}  \bPhiA \bigg] \bigg|_0= -\alc^* \Box \alc +  i\s  \lambda^\dagger\s \frac{\Box}{\PP}\s  \lambda \,.
\label{eq:AbelianL}
\end{align}
It is straightforward to check that this superspace Lagrangian is gauge invariant,
\begin{align}
\label{eq:abelian_gauge_inv}
\mathcal{L} \gauge  \mathcal{L} +  \frac{i^2}{2} \bigg[ \big(\PP_\perp^*\s \bar{\D}\s  \RCA^\dagger\big)\frac{ \Box}{\PP}\s \big(\D\s \bPhiA\big) - \big(\bar{\D}\s \bPhiA^\dagger \big)\s \frac{ \Box}{\PP}\s \big(\PP_\perp\s \D \s\RCA\big) \bigg] \bigg|_0 + \,\, \mathcal{O}\big(\RCA^2\big) = \mathcal{L}\,,
\end{align}
since $\RCA$ is both chiral and anti-chiral.
Note that we have used the fact that $\PP \s \RCA = 0$ to remove any terms arising from the anti-commutator $\{\D, \bar{\D}\} = -2 \s i  \, \PP$.

\subsection{RPI for the Abelian Theory}

As is clear from \Tab{table:RPI-components}, the components of $\bPhiA$ have the same RPI-I and RPI-III transformations as the matter chiral superfield (assuming light-cone gauge).
Therefore, the RPI-I and RPI-III invariance and uniqueness of \Eq{eq:AbelianL} follow from the arguments in \Sec{subsec:chiral_kinetic}.
However, $\bPhiA$ has non-trivial RPI-II transformations, so we must check that \Eq{eq:AbelianL} is consistent with this symmetry.
For the $\lambda$ kinetic term of \Eq{eq:AbelianL}, it is RPI-II invariant for the same reasons as for the $u$ kinetic term of \Eq{eq:fermion_check_rpi_ii}.
Checking RPI-II for the $\alc$ kinetic term requires a new argument.

As is the case with all non-covariant gauge choices, light-cone gauge obscures Lorentz invariance, which here manifests by studying the RPI-II transformations.
Note that $\alc$, $n\cdot A$, and $\bar{n} \cdot A$ transform under RPI analogous to $\PP_\perp$, $\tilde{\PP}$, and $\PP$. 
Under RPI-II, the light cone scalar transforms as $\alc \to \alc + \frac{\kappa_{\rm II}}{\sqrt{2}}\, n \cdot A$, and plugging this into \Eq{eq:AbelianL} yields an apparent violation of RPI-II.
The resolution comes from realizing that  $\bar{n} \cdot A$ transforms as $\bar{n} \cdot A \to \bar{n} \cdot A + \sqrt{2}\left(\kappa_{\rm II}\, \alc^* + \kappa_{\rm II}^*\, \alc \right)$.
Thus, it is unsurprising that fixing $\bar{n}\cdot A =0$ obscures RPI-II.

To verify RPI-II, we need to restore the terms in our Lagrangian that depend on $\bar{n}\cdot A$.
From the top down, the Lagrangian can be derived by expanding the full kinetic term on the light cone.
From the bottom up, though, it is also possible to reconstruct the correct operator by only considering the properties of the effective theory.
In particular, there is a unique gauge artifact term that is linear in $\bar{n}\cdot A$, is RPI-III invariant, and transforms under RPI-I into something that still vanishes once light-cone gauge is enforced.
Adding this term to the Lagrangian, yields
\begin{align}
\mathcal{L} \,\,\,  \supset  \,\,\, - \alc^* \Box \alc + \frac{1}{2} \bar{n} \cdot A \Box (n \cdot A) \,,
\end{align} 
such that under an RPI-II transformation
\begin{align}
\mathcal{L} \RPIii & - \left( \alc^* + \frac{\kappa_{\rm II}^*}{\sqrt{2}} \s n \cdot A \right) \Box \left( \alc + \frac{\kappa_{\rm II}}{\sqrt{2}}\s n \cdot A \right) + \frac{1}{2}\left( \bar{n} \cdot A +\sqrt{2} \left( \kappa_{\rm II}\s \alc^* + \kappa_{\rm II}^*\s \alc \right)\right) \Box (n\cdot A) \nonumber\\ 
& = \mathcal{L} + \frac{\kappa_{\rm II}^*}{\sqrt{2}}\s \big( (n\cdot A \Box \alc)  - \alc \Box (n\cdot A)  \big) + \text{h.c} = \mathcal{L}\,,
\end{align}
where we have integrated by parts and set $\bar{n} \cdot A = 0$ to show the final equality holds.
This demonstrates that the collinear superspace Abelian gauge theory respects both RPI and gauge symmetry.

\subsection{Gauge Transformations and Covariant Derivatives}
Now that we have shown how the Abelian theory can be expressed in collinear superspace, we can lift this to non-Abelian theory, which requires the introduction of interactions.
Each gauge field has a corresponding chiral multiplet $\bPhiA^{a}$ labeled by the group index $a$, with corresponding residual gauge transformations $\RCA^a$.
Furthermore, to write down gauge-invariant interactions, we need to covariantize \Eq{eq:AbelianL}.
To this end, we introduce a non-Abelian covariant derivative in superspace,
\begin{align}
\label{eq:gaugecoveraintderivatives}
\nabla_\perp \bPhialc \equiv \PP_\perp \bPhialc - \frac{i}{\sqrt{2}}  \, g \, \big[\bPhiA^\dagger, \bPhialc\big]  \, , \,\, \quad \quad \nabla_\perp^* \bPhialc \equiv  \PP_\perp^* \bPhialc - \frac{i}{\sqrt{2}}  \, g \, \big[\bPhiA , \bPhialc^\dagger \big] \,,
\end{align}
where we are assuming that both operators are acting on a field with the same charge.
Here, we are using the matrix notation $\bPhiA = T^a \s \bPhiA^a$,  where $T^a$ are the adjoint generators of the gauge group defined as $\big(T^a\big)_{bc} = -  i\s f^{abc}$.
In terms of the matrix components, \Eq{eq:gaugecoveraintderivatives} becomes
 \begin{align}
 \label{eq:nablaComp}
&\nabla^{* ab }_\perp = \PP_\perp^* \delta^{ab} + g f^{cab} \bPhiA^{c}  \, , \,\, \quad \quad \nabla^{ab}_\perp = \PP_\perp \delta^{ab} - g f^{cab} \bPhiA^{c  \dagger}.
\end{align} 
Notice that the lowest component of $\nabla_\perp$ is related to the ordinary gauge-covariant derivative $D_\mu$ as 
\begin{equation}
\nabla_\perp \big|_0 = \PP_\perp - i\s g\s T^a \alc^a = \xi\s \sigma^\mu\s \tilde{\xi}^\dagger \left(\partial_\mu - i\s g\s T^a\s A^a_\mu\right) = \xi\s \sigma^\mu\s \tilde{\xi}^\dagger D_\mu.
\end{equation}
The fermionic component of $\nabla_\perp$ involves the gaugino. 

The gauge transformations are now given by
\begin{align}
\label{eq:NonAbelianPhiGaugeTrans}
\bPhialc &\gauge e^{i\s g\s \RCA} \big(\bPhiA + i\s \sqrt{2}\, \PP_\perp^* \big) e^{-i\s g\s \RCA}, 
\end{align}
where we write the residual gauge transformation parameter in matrix form $\RCA = T^a \s \RCA^a$.
In matrix components,  $\bPhi^a \rightarrow \bPhi^a + (\nabla_\perp^*)_{ab} \RCA_b$, so that \Eq{eq:NonAbelianPhiGaugeTrans} becomes
\begin{align}
\alc^a &\gauge \alc^a +  \PP_\perp \omega^a  + g f^{abc} \alc^b \omega^c \,, \\
 \lambda^a &\gauge \lambda^a + g f^{abc} \omega^b \lambda^c,
\end{align}
and similarly for the conjugate fields.
This verifies that \Eq{eq:NonAbelianPhiGaugeTrans} reproduces the expected non-Abelian gauge transformations of the light-cone-projected degrees of freedom.
The gauge transformation of $\nabla_\perp$ follows from that of $\bPhi_\alc$, with
\begin{alignat}{2}
\nabla_\perp \bPhialc &&\gauge e^{i\s g\s \RCA} \Big(\nabla_\perp \bPhiA - \sqrt{2} \, g \,\PP_\perp \PP_\perp^*  \Big) e^{-i\s g\s \RCA} \,, \notag\\[5pt]
\nabla_\perp^* \bPhialc &&\gauge e^{i\s g\s \RCA} \Big(\nabla_\perp^* \bPhiA -\sqrt{2} \,g \,\PP_\perp^* \PP_\perp^* \Big) e^{-i\s g\s \RCA} \,,
\label{eq:nabla_def_nonAbelian}
\end{alignat}
in matrix notation.

One complication with introducing $\nabla_\perp$ and $\nabla_\perp^*$ is related to RPI.
These objects are inert under RPI-III and transform under RPI-I as
\begin{equation}
\label{eq:nablaperp_rpii}
\nabla_\perp  \bPhialc \RPIi  \nabla_\perp  \bPhialc + \kappa_{\rm I}^*\, \PP  \bPhialc \,,
\end{equation}
and similarly for $\nabla_\perp^*$.
However, $\tilde{\PP}$ transforms under RPI-I as $\tilde{\PP} \to \tilde{\PP} + \rpii\, \PP_\perp+\kappa_{\rm I}^*\, \PP^*_\perp$, and the mismatch between $\PP_\perp$ and $\nabla_\perp$ makes it more complicated to verify RPI-I below.

Another way to view this mismatch is that, because $\bar{n}\cdot A$ and $n\cdot A$ are not present in light-cone gauge, there is no way to write gauge-covariant versions of $\PP$ and $\tilde{\PP}$. 
Importantly, $\nabla_\perp$ and $\nabla_\perp^*$ alone are sufficient for writing down gauge-invariant interactions in a pure gauge theory without matter.
With matter, a covariant version of $\tilde{\PP}$ is required, as discussed in \Sec{subsec:chargedmatter}, which will also help to make RPI-I manifest in the companion paper \cite{Cohen:2019gsc}.

\subsection{Non-Abelian Gauge Theory} 

Using \Eq{eq:gaugecoveraintderivatives}, we can now write down a Lagrangian in superspace that is invariant under the residual non-Abelian gauge transformation $\RCA$.
Making the replacement in \Eq{eq:AbelianL}:
 \begin{equation}
 \label{eq:covariantBoxReplacement}
\frac{\Box}{\PP} = \tilde{\PP} - \frac{\PP_\perp \PP_\perp^*}{\PP} \quad \Longrightarrow \quad \tilde{\PP} - \nabla_\perp \frac{1}{\PP} \nabla^*_\perp \, ,
\end{equation} 
and introducing explicit group indices from \Eq{eq:nablaComp}, 
we have the proposed Lagrangian:   
 \begin{align}
 \label{eq:LnonAbelian}
 \mathcal{L} \,\, &= \,\, \frac{i}{2} \D \bar{\D} \Bigg[ \bPhiA^{a \dagger}  \left( \delta^{ac} \s \tilde{\PP}  - \s \nabla^{ab}_\perp \frac{1}{\PP} \nabla^{*bc}_\perp \right) \bPhiA^c \Bigg] \Bigg|_0  \,.
\end{align} 
We now will argue that \Eq{eq:LnonAbelian} is the unique dimension-four Lagrangian allowed by gauge invariance and RPI.
Dropping terms of order $\RCA^2$ and expanding out the covariant derivatives, the residual gauge transformations result in two classes of terms.
The first class of terms vanish following the same logic as \Eq{eq:abelian_gauge_inv}.
Schematically, these looks like
\begin{align}
\label{eq:gaugeTransOfTildePPterm}
\D \bar{\D} \bigg[ \bPhiA^{a} \, \hat{\mathcal{O}} \, \RCA^a  \bigg]  \Bigg|_0  =  \bigg[ ( \D \bPhiA^{a}) \, \hat{\mathcal{O}} \, (\bar{\D} \RCA^a)  \bigg]  \Bigg|_0 + \big(\text{terms} \propto \PP  \, \RCA^a \big) = 0 \,,
\end{align}
and similarly for the conjugate expression, where $\hat{\mathcal{O}}$ is some differential operator involving $\PP_\perp$, $\PP_\perp^*$, and $\tilde{\PP}$.
Since $\RCA$ is simultaneously chiral and anti-chiral, these vanishes under $\D\s \bar{\D}$.
Additionally, we have invoked the anti-commutation relation $\big\{\D, \bar{\D} \big\} \RCA= 2\s i\, \PP\s \RCA  = 0$.
The second class of terms contain products of fields like $\bPhi_{\alc}^{\dagger a} \, \bPhi_{\alc}^b \, \RCA^c$ with insertions of derivatives and an overall structure constant.
After using integration by parts and combining with Hermitian conjugates terms, these cancel among themselves due to the asymmetry of $f^{abc}$.

To see why RPI-III holds, note that all of the terms in \Eq{eq:LnonAbelian} have the same RPI-III charges as their Abelian counterparts.
RPI-I is a bit more subtle to verify, for the reasons mentioned around \Eq{eq:nablaperp_rpii}.
It is convenient to write out the Lagrangian more explicitly:  
\begin{align}
\label{eq:nonablianLExpanded}
\mathcal{L}
&= \frac{i}{2}\D \bar{\D} \bigg[ \bPhiA^{a \dagger} \frac{\Box}{\PP} \bPhiA^a  \bigg] \bigg|_0 - \frac{i}{2}\s g\s f^{abc}\s \D\s \bar{\D} \bigg[  \left(\bPhiA^{\dagger a}\s \bPhiA^b\right) \frac{\PP_\perp^*}{\PP} \bPhiA^c - \bPhiA^{\dagger a}\s \frac{\PP_\perp}{\PP}\s \left(\bPhiA^{\dagger b}\s \bPhiA^c  \right) \bigg] \bigg|_0 \notag \\[5pt] 
& \qquad - \frac{i}{2} \s g^2\s f^{abc}\s f^{che}\s \D \s \bar{\D} \bigg[  \left(\bPhiA^{\dagger a}\s \bPhiA^b\right) \frac{1}{\PP} \left( \bPhiA^{\dagger h}\s \bPhiA^e \right) \bigg] \bigg|_0  \,\,.
\end{align}
The first and last terms in this expression are manifestly RPI-I invariant.
For the two middle terms, note that $\PP^*_\perp / \PP \to \PP^*_\perp / \PP + \rpii$, so under RPI-I, we have 
\begin{equation}
\label{eq:gauge_rpii_check}
\mathcal{L} \RPIi \mathcal{L}  - 2 \, i \, \rpii  \, \s g\s f^{abc}\s \D\s \bar{\D} \Big[ \bPhiA^{\dagger a}\s \bPhiA^b \bPhiA^c \Big]  + \text{h.c.} = \mathcal{L},
\end{equation}
where in the last step we have used the fact that $f^{abc}$ is completely antisymmetric.
This highlights the link between light-cone gauge invariance and RPI, which can be traced to the link between gauge redundancy and Lorentz invariance.

Verifying RPI-II again requires going to components and checking the invariance explicitly.
Because this is a rather tedious exercise, here we appeal to the top-down construction in~\Refs{Cohen:2016jzp,Cohen:2016dcl}, which had to satisfy RPI-II since it was derived by matching to  the full Lorentz-invariant theory in light-cone gauge.
The expression in \Eq{eq:nonablianLExpanded} is identical to the Lagrangian derived in~\Refs{Cohen:2016jzp,Cohen:2016dcl} (up to conventions and superspace derivative manipulations), which implies that RPI-II is indeed satisfied.
Finally, to see why this term is unique, we can appeal to the same logic as in \Sec{subsec:chiral_kinetic}.
Apart from the replacement of $\PP_\perp$ with $\nabla_\perp$ (and the corresponding replacement of $\Box$ in \Eq{eq:covariantBoxReplacement}), there are no additional ingredients in the gauge case compared to the free chiral multiplet.
Therefore, without introducing any new mass scales, \Eq{eq:LnonAbelian} is the unique Lagrangian one can write consistent with gauge invariance and RPI.

\subsection{Where is Charged Matter?}
\label{subsec:chargedmatter}

Armed with the covariant derivatives in \Eq{eq:gaugecoveraintderivatives}, one might naively think that it would be straightforward to add interactions involving charged matter.
For a charged matter chiral multiplet $\bM$, it should transform under the residual gauge transformation $\RCA$ as
\begin{align}
\bM  &\gauge  e^{i\s g\s \RCA} \bM,\\
\bM^\dagger   &\gauge   \bM^\dagger e^{-i\s g\s \RCA^\dagger} = \bM^\dagger e^{-i\s g\s \RCA},
\end{align}
where we have enforced the chirality/reality of $\RCA = \RCA^\dagger$. 
One can verify that the covariant derivative acts as expected,
\begin{align}
\nabla_\perp^* \bM  \gauge  e^{i\s g\s \RCA} \nabla_\perp^* \bM,
\end{align}
and one might be tempted to propose the candidate Lagrangian:
\begin{equation}
\label{eq:candidatematter}
\mathcal{L}_{\text{candidate}} \,\,\,  \stackrel{?}{\supset} \,\,\, \frac{i}{2} \D \bar{\D} \Bigl[ \bM^\dagger  \left( \tilde{\PP}  - \nabla_\perp \frac{1}{\PP} \nabla_\perp^* \right)  \bM \Bigr] \Big|_{0}  \,,
\end{equation}
in analogy with the non-Abelian gauge kinetic term in \Eq{eq:LnonAbelian}.

It is easy to check, however, that \Eq{eq:candidatematter} is neither gauge invariant nor RPI-I invariant.
Specifically, the manipulations below \Eq{eq:gaugeTransOfTildePPterm} and in \Eq{eq:gauge_rpii_check} no longer work, because arguments invoking the asymmetry of $f^{abc}$ fail when $\bM$ and $\bPhiA$ are distinct fields.
Because $\PP\s \RCA = 0$, we do not need a covariant version of $\PP$ to achieve gauge invariance, but we do need a covariant version of $\tilde{\PP}$ which involves $n \cdot A$.
This same $n \cdot A$ term is relevant for restoring RPI-I, but as discussed many times, this field does not appear in the present light-cone-gauge construction.

Another way to understand why this construction fails is that SUSY gauge theories with charged matter involve non-zero $D$-term auxiliary fields.
In the companion paper, we introduce a novel real superfield with non-trivial RPI transformation properties whose components include $n \cdot A$ and $D$~\cite{Cohen:2019gsc}.
In this way, the restoration of gauge invariance, RPI-I, and the required auxiliary fields are all achieved using related machinery.
From this perspective, the reason why $n \cdot A$ did not need to appear in the pure gauge Lagrangian in \Eq{eq:LnonAbelian} is that $D$ is identically zero in the full $\mathcal{N} = 1$ construction.
We leave a more detailed discussion of this point to the companion paper.

\section{Future Horizons}
\label{sec:Outlook}
In this paper, we provided a set of rules for constructing on-shell SUSY Lagrangians directly in collinear superspace, without any reference to the original Lorentz-invariant description.
This can be contrasted to the approach advocated in~\Refs{Cohen:2016jzp, Cohen:2016dcl}, where the Lagrangian was derived from the full $\mathcal{N} = 1$ theory by fixing a light cone and integrating out non-propagating degrees of freedom in superspace.
We now have a set of fully-consistent EFT rules for collinear superspace, based on the simple restriction given in \Eq{eq:tilde_eta_zero}, which yields a superspace where $\theta^2 = 0$.
This restriction selects a SUSY sub-algebra that is expressed on a light cone, and whose representations are built using only propagating degrees of freedom.
Furthermore, we were able to express the residual light-cone gauge invariance (encoded using the novel superfield $\RCA$), which was then used to derive the Lagrangian of both Abelian and non-Abelian gauge theories.
A formalism for reintroducing non-propagating degrees of freedom will be provided in \Ref{Cohen:2019gsc}, which is necessary to construct Wess-Zumino models and matter/gauge interactions where auxiliary $F$ and $D$ terms are essential.

While this was in some ways an academic exercise for $\mathcal{N}=1$ SUSY, which of course has a simple Lorentz-invariant superspace formulation using off-shell degrees of freedom, there are a number of aspects of our construction which are interesting in their own right.
We described RPI in the language of spinor/helicity, which is not so commonly encountered in the EFT literature. 
We introduced a superspace gauge-covariant derivative $\nabla_\perp$, which has no analog (to our knowledge) in the standard $\mathcal{N}=1$ treatment.
Beyond the novel real and chiral gauge parameter $\RCA$, even more exotic superfields will be encountered in \Ref{Cohen:2019gsc}, defined by mixed constraints involving both spacetime and superspace derivatives.

Ultimately, our hope is that these collinear superspace rules will generalize in a straightforward way to theories with $\mathcal{N} > 1$ SUSY (or even to theories with $d > 4$).
It is well known that the standard superspace approach only works for $\mathcal{N} = 1$, so discovering the underlying rules for an $\mathcal{N} > 1$ collinear superspace could in principle be useful to achieve a deeper understanding of these theories.
For example, perhaps the uniqueness of the $\mathcal{N}=4$ Lagrangian could be proven within collinear superspace directly, or maybe these constructions would illuminate the equivalence of the $\mathcal{N} = 3$ and $\mathcal{N} = 4$ Yang-Mills actions.
Additionally, it would be interesting to search for connections between the collinear superspace formalism and the on-shell recursive approach to scattering amplitudes, see \emph{e.g.}~\cite{Elvang:2013cua} for a review.
By obscuring Lorentz invariance, we hope this formalism will shed additional light on some of the amazing structures that emerge in SUSY field theories.

\acknowledgments

We thank Martin Beneke for a helpful discussion about RPI.
TC is supported by the U.S. Department of Energy, under grant number DE-SC0011640 and DE-SC0018191, and a National Science Foundation LHC Theory Initiative Postdoctoral Fellowship, under grant number PHY-0969510.
GE is supported by the U.S. Department of Energy, under grant numbers DE-SC0011637 and DE-SC0018191, and a National Science Foundation LHC Theory Initiative Postdoctoral Fellowship, under grant number PHY-0969510.
JT thanks the Harvard Center for the Fundamental Laws of Nature for hospitality while this work was completed.
JT is supported by the Office of High Energy Physics of the U.S. Department of Energy under grant DE-SC-0012567 and by the Simons Foundation through a Simons Fellowship in Theoretical Physics.

\appendix
\section{The Generators of RPI}
\label{app:RPIgen}
In this appendix, we discuss the details of the RPI generators~\cite{Kogut:1969xa, Manohar:2002fd}.%
\footnote{See \Ref{Heinonen:2012km} for an analysis of Lorentz invariance and RPI generators in heavy particle effective theories.}
The Poincar\'e group is defined by
\begin{align}
\label{eq:poincare1}
\Big[P_\mu, P_\nu \Big] &= 0\,,\\[5pt] 
\label{eq:poincare2}
\Big[M^{\mu\nu},P^\rho \Big] &= i\s g^{\mu\rho}\,P^\nu-i\,g^{\nu\rho}\,P^\mu \,, \\[5pt]
\label{eq:poincare3}
\Big[M^{\mu \nu}, M^{\kappa \rho}\Big] &= - i g^{\mu \kappa} M^{\nu \rho} - i g^{\nu \rho} M^{\mu \kappa} + i g^{\mu \rho} M^{\nu \kappa} + i g^{\nu \kappa}M^{\mu \rho}\, ,
\end{align}
where $P_\mu = i\s  \partial_\mu$ is the generator of translations, and $M^{\mu \nu}$ is the usual anti-symmetric matrix of Poincar\'e generators
\begin{align}
& M^{\mu \nu} = 
\left[ \begin{array}{cccc}
0 & K^1 & K^2 & K^3  \\
-K^1 & 0 & - J^3 & J^2 \\
-K^2 & J^3 & 0 & -J^1 \\
-K^3 & - J^2 & J^1 & 0
  \end{array} \right],  \,
\end{align}
composed of rotations $M^{ij} = - \epsilon_{ijk}\s J^k$ and boosts $M^{0i} = K^i$, which satisfy the algebra
\begin{align}
\Big[J_i , J_j \Big] = i \s\epsilon_{ijk}\s J_k,\quad \Big[ J_i, K_j \Big] = i \s \epsilon_{ijk} \s K_k, \quad \text{and}\quad \Big[ K_i, K_j\Big] = - i\s \epsilon_{ijk} \s J_k\,. 
\end{align}
Projecting $M^{\mu\nu}$ onto the canonical frame $n^\mu = (1,0,0,1)$ and $\bar{n}^\mu = (1,0,0,-1)$ yields
\begin{align}
\label{eq:LCgenbreaking}
&R_{\rm I}^{\nu_\perp} = \bar{n}_\mu\s M^{\mu \nu_\perp} = M^{0 \nu_\perp} + M^{3 \nu_\perp} \,,\\[3pt] \nonumber
&R_{\rm II}^{\nu_\perp} = n_\mu\s M^{\mu \nu_\perp} = M^{0 \nu_\perp} - M^{3 \nu_\perp}\,, \\[3pt] \nonumber
& R_{\rm III} = n_\mu\s \bar{n}_\nu\s M^{\mu \nu} = 2\, M^{03} = 2\, K^3\,,
\end{align}
where $\nu_\perp = 1,2$, yielding five broken Lorentz generators.  Note that these projections can be equivalently expressed in terms of $\xi$ and $\tilde{\xi}$ using \Eq{eq:twospinorsn_alt}.

We can immediately identify RPI-III as the scalar operator corresponding to boosting along the light cone direction, $\hat{z}$ in the canonical frame.
The remaining four generators 
\begin{align}
R_{\rm I}^{1} = K^1- J^2 \, ,
& \quad R_{\rm I}^{2} = K^2 + J^1\,,\notag \\[3pt] 
R_{\rm II}^{1}  = K^1 + J^2 \, ,&\quad
R_{\rm II}^{2} = K^2 - J^1 \,,
\label{eq:RPI1And2}
\end{align}
correspond to boots and rotations about the directions transverse to the light cone.

By inspecting \Eqs{eq:LCgenbreaking}{eq:RPI1And2}, we see that there is no explicit dependence on the $J_3$ generator.  
This is to be expected since we have picked the canonical frame, which points in the $\hat{z}$-direction, and the RPI transformations are the combinations of rotations and boosts which leave this direction unchanged.
However, when we compute the commutators
\begin{align}
\begin{array}{ll}
  \Big[ R^{\s\mu_\perp}_{\rm I}, R^{\nu_\perp}_{\rm I} \Big] = 0\,, &\qquad\qquad\qquad   \Big[ R^{\s\mu_\perp}_{\rm II}, R^{\nu_\perp}_{\rm II} \Big] = 0\,, \\[12pt]
 \Big[ R_{\rm I}^1, R_{\rm II}^1 \Big] =  i\s R_{\rm III}\,, &\qquad\qquad\qquad  \Big[ R_{\rm I}^2, R_{\rm II}^2 \Big] = i\s R_{\rm III}\,, \\[12pt]
  \Big[ R_{\rm I}^1, R_{\rm II}^2 \Big] =  -2\s i \s J_3\,, &\qquad\qquad\qquad  \Big[ R_{\rm I}^2, R_{\rm II}^2 \Big] = 2\s i\s J_3\,, \\[12pt]
  \Big[ R_{\rm I}^1, R_{\rm III} \Big] = -2\s i\s R_{\rm I}^1\,, &\qquad\qquad\qquad  \Big[ R_{\rm I}^2, R_{\rm III} \Big] = -2\s i\s R_{\rm I}^2\,, \\[12pt]
    \Big[ R_{\rm II}^1, R_{\rm III} \Big] = 2\s i\s R_{\rm II}^1\,, &\qquad\qquad\qquad  \Big[ R_{\rm II}^2, R_{\rm III} \Big] = 2\s i\s R_{\rm II}^2 \,,
\end{array}
  \label{eq:RPIComm}
\end{align}
we see $J_3$ is generated by successive RPI transformations.  
This is exactly the sense in which RPI secretly encodes the full Lorentz invariance, in that one can reconstruct the ``missing'' $J_3$ generator through the application of the RPI transformations alone.

\end{spacing}

\section{The Rigging:  A Summary of Useful Formulae}
\label{app:UsefulFormulas}

This appendix provides a set of reference formulas that are useful for deriving the results presented in the main text.  
Note that some expressions are redundant with the body of the paper, but we reproduce them here for convenience.

We work in Minkowski space with metric signature $g^{\mu \nu} = \textrm{diag}\left(+1,-1,-1,-1\right)$, and our $\gamma$-matrices are in the Weyl basis.
We follow spinor conventions of \Refs{Dreiner:2008tw, Binetruy:2006ad}. 
For a useful review of the conventions relevant for SUSY see pages 449--453 of \Ref{Binetruy:2006ad}.

\subsection{Frame-Independent Expressions}

Here, we briefly summarize the frame-independent expressions.
For light-cone derivative we have:
\begin{align}
\bar{\sigma}^\mu \partial_\mu 
&= \tilde{\xi}^{\dagger \dot{\alpha}} \tilde{\xi}^\alpha\, \PP + \xi^{\dagger \dot{\alpha}} \xi^\alpha\, \tilde{\PP} + \xi^{\dagger \dot{\alpha}}  \tilde{\xi}^\alpha\, \PP_\perp + \tilde{\xi}^{\dagger \dot{\alpha}}  \xi^\alpha\, \PP^*_\perp \,, \notag\\[4pt]
\sigma^\mu \partial_\mu &=  \tilde{\xi}_\alpha \tilde{\xi}^\dagger_{\dot{\alpha}}\, \PP + \xi_\alpha \xi^\dagger_{\dot{\alpha}} \,\tilde{\PP} + \xi_\alpha \tilde{\xi}^\dagger_{\dot{\alpha}}\, \PP_\perp^* + \tilde{\xi}_\alpha \xi^\dagger_{\dot{\alpha}}\, \PP_\perp\, .
\label{eq:sigmapartialdecom}
\end{align}
Note that we can write $\Box = \PP \s \tilde{\PP} - \PP_\perp^* \PP_\perp$.
For light-cone spinors we have:
\bea
u_\alpha = \tilde{\xi}_\alpha u - \xi_\alpha \uu \qquad \text{so that} \qquad u = \xi^\alpha u_\alpha \,, \qquad \text{and} \qquad \tilde{\xi}^\alpha u_\alpha = \uu\,,
\label{eq:fermionDecomp}
\eea
where the choice of convention will be discussed further below.
Finally for the gauge field we have:
\begin{align}
\left(\sigma \cdot A \right)_{\alpha \dot{\alpha}} = \xi_\alpha\, \xi^{\dagger}_{\dot{\alpha}}\, n\cdot A + \tilde{\xi}_\alpha\, \tilde{\xi}^\dagger_{\dot{\alpha}}\, \bar{n}\cdot A+\sqrt{2} \,  \xi_\alpha\, \tilde{\xi}^\dagger_{\dot{\alpha}}\, \alc^* +\sqrt{2} \,  \tilde{\xi}_\alpha\, \xi^{\dagger}_{\dot{\alpha}}\, \alc\,.
\end{align}

\subsection{The Canonical Frame}
We often appeal to the \emph{canonical frame}, which is specified by 
\begin{align}
n^\mu = (1,0,0,1)\,, \qquad \bar{n}^\mu = (1,0,0,-1)\,,
\end{align}
and can be rewritten using a spinor helicity decomposition as
\begin{align}
n^\mu = \tilde{\xi}^\dag \bar{\sigma}^\mu \tilde{\xi} = \tilde{\xi} \sigma^\mu \tilde{\xi}^\dag \, \qquad \text{and} \qquad \bar{n}^\mu = \xi^\dag \bar{\sigma}^\mu \xi = \xi \sigma^\mu \xi^\dag\,.
\end{align}
This is equivalent to fixing the spinors to
\begin{align}
\label{eq:bosoniccanonicalspinors_again}
\xi^\alpha = (0,1)\,,\qquad \xi_\alpha = (-1,0)^{\intercal}, \qquad \tilde{\xi}^\alpha = (1,0)\,, \qquad \tilde{\xi}_\alpha = (0,1)^{\intercal}\,,
\end{align}
where
\begin{equation}
\xi^\alpha\s \tilde{\xi}_\alpha = 1, \qquad \tilde{\xi}^\alpha\s \xi_\alpha = - \epsilon^{\alpha \beta}\, \xi_\alpha \s\tilde{\xi}_\beta =-1\,, \qquad\xi^\dagger_{\dot{\alpha}}\s \tilde{\xi}^{\dagger \dot{\alpha}} = -1, \qquad \tilde{\xi}^\dag_{\dot{\alpha}}\, \xi^{\dag \dot{\alpha}} = 1\,.
\end{equation}

To express a Weyl spinor fermion in the canonical frame, we note that the projection operators that act as
\begin{align}
P_n u_\alpha = \frac{n\cdot\sigma}{2} \frac{\bar{n}\cdot \bar{\sigma}}{2}\, u_\alpha = u_2 \qquad \text{and} \qquad P_{\bar{n}} u_\alpha =\frac{\bar{n}\cdot\sigma}{2} \frac{n\cdot \bar{\sigma}}{2}\, u_\alpha =  u_1,
\end{align}
where we note that the $\alpha$ index is lowered in these expressions. 
We identify $u_2$ as the helicity aligned with the light-cone and $u_1$ is anti-aligned helicity, so that
\begin{align}
u \equiv u_2  \qquad \text{and} \qquad \tilde{u} \equiv u_1\,.
\end{align}
Given \eqref{eq:bosoniccanonicalspinors_again} this can be all be made consistent with the following conventional choice: 
\begin{align}
&u_\alpha =  \tilde{\xi}_\alpha u - \xi_\alpha \tilde{u}\,  \qquad \text{so that} \qquad \xi^\alpha u_\alpha = u \quad \text{and} \quad \tilde{\xi}^\alpha u_\alpha = \tilde{u}\,.
\end{align}
We note that this minus sign is not required for the expansion of $\theta^\alpha$ on the lightcone, and so we do not include it, see \eqref{eq:eta_def}.

A vector $V^\mu$ can be decomposed on the light cone as
\begin{align}
n\cdot V &= \tilde{\xi}^\dag \bar{\sigma}\cdot V \tilde{\xi} = \tilde{\xi} \sigma\cdot V \tilde{\xi}^\dag= V_0 + V_3 = -\big(V^0 + V^3\big)\,, \notag \\[5pt]
\bar{n} \cdot V &= \xi^\dag \bar{\sigma}\cdot V \xi = \xi \sigma\cdot V \xi^\dag= V_0 - V_3 = -\big(V^0 + V^3\big)\,,\notag\\[5pt]
V_\perp &= \tilde{\xi}^\dag \bar{\sigma}\cdot V \xi = \xi \sigma\cdot V \tilde{\xi}^\dag =V_1 + i\s V_2 = -\big(V^1 + i\s V^2\big)\,, \notag\\[5pt] 
V_\perp^{*} &= \xi^\dag \bar{\sigma}\cdot V \tilde{\xi} = \tilde{\xi} \sigma\cdot V \xi^\dag =V_1 - i\s V_2 = -\big(V^1 - i\s V^2\big)\,, 
\end{align}
where in the final steps we have fixed to the canonical frame.
This can be conveniently packaged as
\begin{align}
\sigma \cdot V = 
\left(\begin{matrix}
n\cdot V & -V_\perp^* \\
-V_\perp & \bar{n}\cdot V
\end{matrix}\right)
\qquad\text{and} \qquad
\bar{\sigma} \cdot V = 
\left(\begin{matrix}
\bar{n}\cdot V & V_\perp \\
V_\perp^* & n\cdot V
\end{matrix}\right)\,.
\end{align}

\subsection{Conventions in Soft-Collinear SUSY}

It is useful to keep in mind the spinor structure of objects in LCG used in the soft-collinear SUSY paper \cite{Cohen:2016dcl}.
For instance,
\begin{align}
& \sigma^\mu \partial_\mu = 
\left[ \begin{array}{cc}
\np & \sqrt{2}\,\partial^*  \\
\sqrt{2}\, \partial & \nbp  \end{array} \right]_{\alpha \dot{\alpha}},  \,
& \bar{\sigma}^\mu \partial_\mu = 
\left[ \begin{array}{cc}
\nbp & -\sqrt{2}\,\partial^*  \\
-\sqrt{2}\, \partial & \np  \end{array} \right]^{\dot{\alpha} \alpha}.
\end{align}
Note the $\sqrt{2}$ difference between the $\partial_\perp$ definitions and the $\PP_\perp$ definition used in this paper.
Similar expressions hold for other contractions such as $\sigma^\mu A_\mu$.
These expressions are independent of the choice of $n^\mu$ and $\bar{n}^\mu$ direction. 

Note that throughout we include the Lorentz contraction in the definitions of $\partial_\perp^2$, as this is convenient when working with LCG scalars:
\bea
\partial_\perp^2 \equiv \partial_\perp^\mu \partial_{\perp \mu}  = - \partial_1^2 - \partial_2^2  = - 2\, \partial\s \partial^* \, ,
\eea
where we have converted to LCG derivatives.
This is in contrast to some places in the literature which relate $\partial_\perp^2$ to the explicit component expression with the opposite sign.
In terms of this notation 
\bea
\Box = \partial^\mu\s \partial_\mu = \nbp \,\np + \partial_\perp^2 = \nbp \s\np - 2\, \partial\s \partial^* \,.
\eea

\begin{spacing}{1.1}
\addcontentsline{toc}{section}{\protect\numberline{}References}%
\bibliography{OnShellSUSY}

\providecommand{\href}[2]{#2}\begingroup\raggedright\begin{thebibliography}{10}

\bibitem{Seiberg:1994pq}
N.~Seiberg, ``{Electric - magnetic duality in supersymmetric nonAbelian gauge
  theories},'' \href{http://dx.doi.org/10.1016/0550-3213(94)00023-8}{{\em Nucl.
  Phys.} {\bf B435} (1995)  129--146},
\href{http://arxiv.org/abs/hep-th/9411149}{{\tt arXiv:hep-th/9411149
  [hep-th]}}.

\bibitem{Seiberg:1994aj}
N.~Seiberg and E.~Witten, ``{Monopoles, duality and chiral symmetry breaking in
  N=2 supersymmetric QCD},''
  \href{http://dx.doi.org/10.1016/0550-3213(94)90214-3}{{\em Nucl. Phys.} {\bf
  B431} (1994)  484--550},
\href{http://arxiv.org/abs/hep-th/9408099}{{\tt arXiv:hep-th/9408099
  [hep-th]}}.

\bibitem{Seiberg:1994rs}
N.~Seiberg and E.~Witten, ``{Electric - magnetic duality, monopole
  condensation, and confinement in N=2 supersymmetric Yang-Mills theory},''
  \href{http://dx.doi.org/10.1016/0550-3213(94)90124-4}{{\em Nucl. Phys.} {\bf
  B426} (1994)  19--52}, \href{http://arxiv.org/abs/hep-th/9407087}{{\tt
  arXiv:hep-th/9407087 [hep-th]}}.
[Erratum: Nucl. Phys.B430,485(1994)].

\bibitem{Maldacena:1997re}
J.~M. Maldacena, ``{The Large N limit of superconformal field theories and
  supergravity},'' \href{http://dx.doi.org/10.1023/A:1026654312961}{{\em Int.
  J. Theor. Phys.} {\bf 38} (1999)  1113--1133},
  \href{http://arxiv.org/abs/hep-th/9711200}{{\tt arXiv:hep-th/9711200
  [hep-th]}}.
[Adv. Theor. Math. Phys.2,231(1998)].

\bibitem{Pestun:2007rz}
V.~Pestun, ``{Localization of gauge theory on a four-sphere and supersymmetric
  Wilson loops},'' \href{http://dx.doi.org/10.1007/s00220-012-1485-0}{{\em
  Commun. Math. Phys.} {\bf 313} (2012)  71--129},
\href{http://arxiv.org/abs/0712.2824}{{\tt arXiv:0712.2824 [hep-th]}}.

\bibitem{Erickson:2000af}
J.~K. Erickson, G.~W. Semenoff, and K.~Zarembo, ``{Wilson loops in N=4
  supersymmetric Yang-Mills theory},''
  \href{http://dx.doi.org/10.1016/S0550-3213(00)00300-X}{{\em Nucl. Phys.} {\bf
  B582} (2000)  155--175},
\href{http://arxiv.org/abs/hep-th/0003055}{{\tt arXiv:hep-th/0003055
  [hep-th]}}.

\bibitem{Bianchi:2008pu}
M.~Bianchi, H.~Elvang, and D.~Z. Freedman, ``{Generating Tree Amplitudes in N=4
  SYM and N = 8 SG},''
  \href{http://dx.doi.org/10.1088/1126-6708/2008/09/063}{{\em JHEP} {\bf 09}
  (2008)  063},
\href{http://arxiv.org/abs/0805.0757}{{\tt arXiv:0805.0757 [hep-th]}}.

\bibitem{Komargodski:2009rz}
Z.~Komargodski and N.~Seiberg, ``{From Linear SUSY to Constrained
  Superfields},'' \href{http://dx.doi.org/10.1088/1126-6708/2009/09/066}{{\em
  JHEP} {\bf 09} (2009)  066},
\href{http://arxiv.org/abs/0907.2441}{{\tt arXiv:0907.2441 [hep-th]}}.

\bibitem{Festuccia:2011ws}
G.~Festuccia and N.~Seiberg, ``{Rigid Supersymmetric Theories in Curved
  Superspace},'' \href{http://dx.doi.org/10.1007/JHEP06(2011)114}{{\em JHEP}
  {\bf 06} (2011)  114},
\href{http://arxiv.org/abs/1105.0689}{{\tt arXiv:1105.0689 [hep-th]}}.

\bibitem{Kahn:2015mla}
Y.~Kahn, D.~A. Roberts, and J.~Thaler, ``{The goldstone and goldstino of
  supersymmetric inflation},''
  \href{http://dx.doi.org/10.1007/JHEP10(2015)001}{{\em JHEP} {\bf 10} (2015)
  001},
\href{http://arxiv.org/abs/1504.05958}{{\tt arXiv:1504.05958 [hep-th]}}.

\bibitem{Ferrara:2015tyn}
S.~Ferrara, R.~Kallosh, and J.~Thaler, ``{Cosmology with orthogonal nilpotent
  superfields},'' \href{http://dx.doi.org/10.1103/PhysRevD.93.043516}{{\em
  Phys. Rev.} {\bf D93} (2016) no.~4, 043516},
\href{http://arxiv.org/abs/1512.00545}{{\tt arXiv:1512.00545 [hep-th]}}.

\bibitem{Dall'Agata:2016yof}
G.~Dall'Agata, E.~Dudas, and F.~Farakos, ``{On the origin of constrained
  superfields},''
\href{http://arxiv.org/abs/1603.03416}{{\tt arXiv:1603.03416 [hep-th]}}.

\bibitem{Delacretaz:2016nhw}
L.~V. Delacretaz, V.~Gorbenko, and L.~Senatore, ``{The Supersymmetric Effective
  Field Theory of Inflation},''
  \href{http://dx.doi.org/10.1007/JHEP03(2017)063}{{\em JHEP} {\bf 03} (2017)
  063},
\href{http://arxiv.org/abs/1610.04227}{{\tt arXiv:1610.04227 [hep-th]}}.

\bibitem{Cacciatori:2017qyd}
S.~L. Cacciatori and S.~Noja, ``{Projective Superspaces in Practice},''
  \href{http://dx.doi.org/10.1016/j.geomphys.2018.03.021}{{\em J. Geom. Phys.}
  {\bf 130} (2018)  40--62},
\href{http://arxiv.org/abs/1708.02820}{{\tt arXiv:1708.02820 [math.AG]}}.

\bibitem{Salam:1974yz}
A.~Salam and J.~A. Strathdee, ``{Supergauge Transformations},''
\href{http://dx.doi.org/10.1016/0550-3213(74)90537-9}{{\em Nucl. Phys.} {\bf
  B76} (1974)  477--482}.

\bibitem{Ferrara:1974ac}
S.~Ferrara, J.~Wess, and B.~Zumino, ``{Supergauge Multiplets and
  Superfields},''
\href{http://dx.doi.org/10.1016/0370-2693(74)90283-4}{{\em Phys. Lett.} {\bf
  51B} (1974)  239}.

\bibitem{Galperin:2001uw}
A.~S. Galperin, E.~A. Ivanov, V.~I. Ogievetsky, and E.~S. Sokatchev,
  \href{http://dx.doi.org/10.1017/CBO9780511535109}{{\em {Harmonic
  superspace}}}.
\newblock Cambridge Monographs on Mathematical Physics. Cambridge University
  Press, 2007.
\newblock
\url{http://www.cambridge.org/mw/academic/subjects/physics/theoretical-physics-and-mathematical-physics/harmonic-superspace?format=PB}.
\newblock

\bibitem{Galperin:1985ec}
A.~Galperin, E.~Ivanov, V.~Ogievetsky, and E.~Sokatchev, ``{HARMONIC
  SUPERSPACE: KEY TO N=2 SUPERSYMMETRY THEORIES},'' {\em JETP Lett.} {\bf 40}
  (1984)  912--916.
[,418(1985)].

\bibitem{Akulov:1988tm}
V.~P. Akulov, D.~P. Sorokin, and I.~A. Bandos, ``{Particle Mechanics in
  Harmonic Superspace},''
\href{http://dx.doi.org/10.1142/S0217732388001951}{{\em Mod. Phys. Lett.} {\bf
  A3} (1988)  1633--1645}.

\bibitem{Howe:1998jw}
P.~S. Howe, ``{On harmonic superspace},''
  \href{http://dx.doi.org/10.1007/BFb0104588}{{\em Lect. Notes Phys.} {\bf 524}
  (1999)  68}, \href{http://arxiv.org/abs/hep-th/9812133}{{\tt
  arXiv:hep-th/9812133 [hep-th]}}.
[,68(1998)].

\bibitem{Davgadorj:2017ezp}
A.~Davgadorj and R.~von Unge, ``{$\mathcal{N} = 2$ super Yang-Mills theory in
  projective superspace},''
  \href{http://dx.doi.org/10.1103/PhysRevD.97.105017}{{\em Phys. Rev.} {\bf
  D97} (2018) no.~10, 105017},
\href{http://arxiv.org/abs/1706.07000}{{\tt arXiv:1706.07000 [hep-th]}}.

\bibitem{Galperin:1984av}
A.~Galperin, E.~Ivanov, S.~Kalitsyn, V.~Ogievetsky, and E.~Sokatchev,
  ``{Unconstrained N=2 Matter, Yang-Mills and Supergravity Theories in Harmonic
  Superspace},'' \href{http://dx.doi.org/10.1088/0264-9381/1/5/004}{{\em Class.
  Quant. Grav.} {\bf 1} (1984)  469--498}.
[Erratum: Class. Quant. Grav.2,127(1985)].

\bibitem{Sokatchev:1985tc}
E.~Sokatchev, ``{Light Cone Harmonic Superspace and Its Applications},''
\href{http://dx.doi.org/10.1016/0370-2693(86)90652-0}{{\em Phys. Lett.} {\bf
  169B} (1986)  209--214}.

\bibitem{Ivanov:1984ut}
E.~Ivanov, S.~Kalitsyn, A.~V. Nguyen, and V.~Ogievetsky, ``{Harmonic
  Superspaces of Extended Supersymmetry. The Calculus of Harmonic Variables},''
\href{http://dx.doi.org/10.1088/0305-4470/18/17/026}{{\em J. Phys.} {\bf A18}
  (1985)  3433}.

\bibitem{Ohta:1985au}
N.~Ohta and H.~Yamaguchi, ``{Superfield Perturbation Theory in Harmonic
  Superspace},''
\href{http://dx.doi.org/10.1103/PhysRevD.32.1954}{{\em Phys. Rev.} {\bf D32}
  (1985)  1954}.

\bibitem{Ivanov:2003nk}
E.~Ivanov and O.~Lechtenfeld, ``{N=4 supersymmetric mechanics in harmonic
  superspace},'' \href{http://dx.doi.org/10.1088/1126-6708/2003/09/073}{{\em
  JHEP} {\bf 09} (2003)  073},
\href{http://arxiv.org/abs/hep-th/0307111}{{\tt arXiv:hep-th/0307111
  [hep-th]}}.

\bibitem{Mandelstam:1982cb}
S.~Mandelstam, ``{Light Cone Superspace and the Ultraviolet Finiteness of the
  N=4 Model},''
\href{http://dx.doi.org/10.1016/0550-3213(83)90179-7}{{\em Nucl. Phys.} {\bf
  B213} (1983)  149--168}.

\bibitem{Brink:1982pd}
L.~Brink, O.~Lindgren, and B.~E.~W. Nilsson, ``{N=4 Yang-Mills Theory on the
  Light Cone},''
\href{http://dx.doi.org/10.1016/0550-3213(83)90678-8}{{\em Nucl. Phys.} {\bf
  B212} (1983)  401}.

\bibitem{Brink:1982wv}
L.~Brink, O.~Lindgren, and B.~E.~W. Nilsson, ``{The Ultraviolet Finiteness of
  the N=4 Yang-Mills Theory},''
\href{http://dx.doi.org/10.1016/0370-2693(83)91210-8}{{\em Phys. Lett.} {\bf
  123B} (1983)  323--328}.

\bibitem{Cohen:2019gsc}
T.~Cohen, G.~Elor, A.~J. Larkoski, and J.~Thaler, ``{Circumnavigating Collinear
  Superspace},''
\href{http://arxiv.org/abs/1909.00009}{{\tt arXiv:1909.00009 [hep-th]}}.

\bibitem{Cohen:2016jzp}
T.~Cohen, G.~Elor, and A.~J. Larkoski, ``{Collinear Superspace},''
  \href{http://dx.doi.org/10.1103/PhysRevD.93.125013}{{\em Phys. Rev.} {\bf
  D93} (2016) no.~12, 125013},
\href{http://arxiv.org/abs/1603.09346}{{\tt arXiv:1603.09346 [hep-th]}}.

\bibitem{Cohen:2016dcl}
T.~Cohen, G.~Elor, and A.~J. Larkoski, ``{Soft-Collinear Supersymmetry},''
  \href{http://dx.doi.org/10.1007/JHEP03(2017)017}{{\em JHEP} {\bf 03} (2017)
  017},
\href{http://arxiv.org/abs/1609.04430}{{\tt arXiv:1609.04430 [hep-th]}}.

\bibitem{Bauer:2000yr}
C.~W. Bauer, S.~Fleming, D.~Pirjol, and I.~W. Stewart, ``{An Effective field
  theory for collinear and soft gluons: Heavy to light decays},''
  \href{http://dx.doi.org/10.1103/PhysRevD.63.114020}{{\em Phys. Rev.} {\bf
  D63} (2001)  114020},
\href{http://arxiv.org/abs/hep-ph/0011336}{{\tt arXiv:hep-ph/0011336
  [hep-ph]}}.

\bibitem{Bauer:2001ct}
C.~W. Bauer and I.~W. Stewart, ``{Invariant operators in collinear effective
  theory},'' \href{http://dx.doi.org/10.1016/S0370-2693(01)00902-9}{{\em Phys.
  Lett.} {\bf B516} (2001)  134--142},
\href{http://arxiv.org/abs/hep-ph/0107001}{{\tt arXiv:hep-ph/0107001
  [hep-ph]}}.

\bibitem{Bauer:2001yt}
C.~W. Bauer, D.~Pirjol, and I.~W. Stewart, ``{Soft collinear factorization in
  effective field theory},''
  \href{http://dx.doi.org/10.1103/PhysRevD.65.054022}{{\em Phys. Rev.} {\bf
  D65} (2002)  054022},
\href{http://arxiv.org/abs/hep-ph/0109045}{{\tt arXiv:hep-ph/0109045
  [hep-ph]}}.

\bibitem{Belitsky:2004yg}
A.~V. Belitsky, S.~E. Derkachov, G.~P. Korchemsky, and A.~N. Manashov,
  ``{Quantum integrability in superYang-Mills theory on the light cone},''
  \href{http://dx.doi.org/10.1016/j.physletb.2004.04.092}{{\em Phys. Lett.}
  {\bf B594} (2004)  385--401},
\href{http://arxiv.org/abs/hep-th/0403085}{{\tt arXiv:hep-th/0403085
  [hep-th]}}.

\bibitem{Kallosh:2009db}
R.~Kallosh, ``{N=8 Supergravity on the Light Cone},''
  \href{http://dx.doi.org/10.1103/PhysRevD.80.105022}{{\em Phys. Rev.} {\bf
  D80} (2009)  105022},
\href{http://arxiv.org/abs/0903.4630}{{\tt arXiv:0903.4630 [hep-th]}}.

\bibitem{Green:1996um}
M.~B. Green and M.~Gutperle, ``{Light cone supersymmetry and d-branes},''
  \href{http://dx.doi.org/10.1016/0550-3213(96)00352-5}{{\em Nucl. Phys.} {\bf
  B476} (1996)  484--514},
\href{http://arxiv.org/abs/hep-th/9604091}{{\tt arXiv:hep-th/9604091
  [hep-th]}}.

\bibitem{Hearin:2010dw}
P.~Hearin, ``{Light-Cone Superspace BPS Theory},''
  \href{http://dx.doi.org/10.1016/j.nuclphysb.2011.01.005}{{\em Nucl. Phys.}
  {\bf B846} (2011)  226--249},
\href{http://arxiv.org/abs/1008.3877}{{\tt arXiv:1008.3877 [hep-th]}}.

\bibitem{Ramond:2009hb}
P.~Ramond, ``{Still in Light-Cone Superspace},''
  \href{http://dx.doi.org/10.1142/S0217751X10048676}{{\em Int. J. Mod. Phys.}
  {\bf A25} (2010)  367--380},
\href{http://arxiv.org/abs/0910.1993}{{\tt arXiv:0910.1993 [hep-th]}}.

\bibitem{Siegel:1981ec}
W.~Siegel and S.~J. Gates, Jr., ``{SUPERPROJECTORS},''
\href{http://dx.doi.org/10.1016/0550-3213(81)90382-5}{{\em Nucl. Phys.} {\bf
  B189} (1981)  295--316}.

\bibitem{Brink:1981nb}
L.~Brink and J.~H. Schwarz, ``{Quantum Superspace},''
\href{http://dx.doi.org/10.1016/0370-2693(81)90093-9}{{\em Phys. Lett.} {\bf
  100B} (1981)  310--312}.

\bibitem{Manohar:2002fd}
A.~V. Manohar, T.~Mehen, D.~Pirjol, and I.~W. Stewart, ``{Reparameterization
  invariance for collinear operators},''
  \href{http://dx.doi.org/10.1016/S0370-2693(02)02029-4}{{\em Phys. Lett.} {\bf
  B539} (2002)  59--66},
\href{http://arxiv.org/abs/hep-ph/0204229}{{\tt arXiv:hep-ph/0204229
  [hep-ph]}}.

\bibitem{Becher:2014oda}
T.~Becher, A.~Broggio, and A.~Ferroglia, ``{Introduction to Soft-Collinear
  Effective Theory},'' \href{http://dx.doi.org/10.1007/978-3-319-14848-9}{{\em
  Lect. Notes Phys.} {\bf 896} (2015)  pp.1--206},
\href{http://arxiv.org/abs/1410.1892}{{\tt arXiv:1410.1892 [hep-ph]}}.

\bibitem{Gates:1982an}
S.~J. Gates, Jr., ``{On-shell and Conformal $N=4$ Supergravity in
  Superspace},''
\href{http://dx.doi.org/10.1016/0550-3213(83)90229-8}{{\em Nucl. Phys.} {\bf
  B213} (1983)  409--444}.

\bibitem{Brink:1980cb}
L.~Brink, ``{The On-shell $N$=8 Supergravity in Superspace},'' in {\em
  {Proceedings for Unification of the Fundamental Particle Interactions}},
  p.~157.
\newblock
1980.
\newblock

\bibitem{Elvang:2013cua}
H.~Elvang and Y.-t. Huang, ``{Scattering Amplitudes},''
\href{http://arxiv.org/abs/1308.1697}{{\tt arXiv:1308.1697 [hep-th]}}.

\bibitem{Haag:1974qh}
R.~Haag, J.~T. Lopuszanski, and M.~Sohnius, ``{All Possible Generators of
  Supersymmetries of the $S$-Matrix},''
\href{http://dx.doi.org/10.1016/0550-3213(75)90279-5}{{\em Nucl. Phys.} {\bf
  B88} (1975)  257}.

\bibitem{Coleman:1967ad}
S.~R. Coleman and J.~Mandula, ``{All Possible Symmetries of the S Matrix},''
\href{http://dx.doi.org/10.1103/PhysRev.159.1251}{{\em Phys. Rev.} {\bf 159}
  (1967)  1251--1256}.

\bibitem{Dreiner:2008tw}
H.~K. Dreiner, H.~E. Haber, and S.~P. Martin, ``{Two-component spinor
  techniques and Feynman rules for quantum field theory and supersymmetry},''
  \href{http://dx.doi.org/10.1016/j.physrep.2010.05.002}{{\em Phys. Rept.} {\bf
  494} (2010)  1--196},
\href{http://arxiv.org/abs/0812.1594}{{\tt arXiv:0812.1594 [hep-ph]}}.

\bibitem{Binetruy:2006ad}
P.~Binetruy, {\em {Supersymmetry: Theory, experiment and cosmology}}.
\newblock Oxford University Press,
2006.
\newblock

\bibitem{Leibbrandt:1983pj}
G.~Leibbrandt, ``{The Light Cone Gauge in Yang-Mills Theory},''
\href{http://dx.doi.org/10.1103/PhysRevD.29.1699}{{\em Phys. Rev.} {\bf D29}
  (1984)  1699}.

\bibitem{Larkoski:2014bxa}
A.~J. Larkoski, D.~Neill, and I.~W. Stewart, ``{Soft Theorems from Effective
  Field Theory},'' \href{http://dx.doi.org/10.1007/JHEP06(2015)077}{{\em JHEP}
  {\bf 06} (2015)  077},
\href{http://arxiv.org/abs/1412.3108}{{\tt arXiv:1412.3108 [hep-th]}}.

\bibitem{Marcantonini:2008qn}
C.~Marcantonini and I.~W. Stewart, ``{Reparameterization Invariant Collinear
  Operators},'' \href{http://dx.doi.org/10.1103/PhysRevD.79.065028}{{\em Phys.
  Rev.} {\bf D79} (2009)  065028},
\href{http://arxiv.org/abs/0809.1093}{{\tt arXiv:0809.1093 [hep-ph]}}.

\bibitem{Bertolini:2013via}
D.~Bertolini, J.~Thaler, and Z.~Thomas,
  \href{http://dx.doi.org/10.1142/9789814525220_0009}{``{Super-Tricks for
  Superspace},''} pp.~421--496.
\newblock 2013.
\newblock
\href{http://arxiv.org/abs/1302.6229}{{\tt arXiv:1302.6229 [hep-ph]}}.
\newblock

\bibitem{Leibbrandt:1987qv}
G.~Leibbrandt, ``{Introduction to Noncovariant Gauges},''
\href{http://dx.doi.org/10.1103/RevModPhys.59.1067}{{\em Rev. Mod. Phys.} {\bf
  59} (1987)  1067}.

\bibitem{Kogut:1969xa}
J.~B. Kogut and D.~E. Soper, ``{Quantum Electrodynamics in the Infinite
  Momentum Frame},''
\href{http://dx.doi.org/10.1103/PhysRevD.1.2901}{{\em Phys. Rev.} {\bf D1}
  (1970)  2901--2913}.

\bibitem{Heinonen:2012km}
J.~Heinonen, R.~J. Hill, and M.~P. Solon, ``{Lorentz invariance in heavy
  particle effective theories},''
  \href{http://dx.doi.org/10.1103/PhysRevD.86.094020}{{\em Phys. Rev.} {\bf
  D86} (2012)  094020},
\href{http://arxiv.org/abs/1208.0601}{{\tt arXiv:1208.0601 [hep-ph]}}.

\end{thebibliography}\endgroup
\bibliographystyle{utphys}
\end{spacing}

\end{document}